\documentclass[12pt, onecolumn, triplespace,
journal]{IEEEtran}                                      
\usepackage{rotating}
\usepackage{amsmath}
\ifCLASSINFOpdf
\usepackage{graphicx,amssymb}
\usepackage{algorithm}
 \usepackage{algorithmic}
 \usepackage{diagbox}

\makeatletter

\@addtoreset{algorithm}{section}
\makeatother

\usepackage{pdfsync}
\usepackage{amsthm}

\newtheorem{theorem}{Theorem}[section]

\newtheorem{proposition}[theorem]{Proposition}
\newtheorem{definition}[theorem]{Definition}
\newtheorem{corollary}[theorem]{Corollary}

\newtheorem{remark}[theorem]{Remark}

\newtheorem{assumption}[theorem]{Assumption}

\def\CC{ {\mathbb C} }

\numberwithin{equation}{section}
\usepackage{color}

\begin{document}


%

\title{Nonsubsampled Graph Filter Banks and Distributed Implementation} 
\author{Junzheng Jiang, Cheng Cheng,  and Qiyu Sun 
\thanks{Jiang is with the School of Information and Communication, Guilin University of Electronic Technology, Guilin 541004, China;
\and Cheng is with the Department of Mathematics, Duke University,  and Statistical and Applied Mathematical Sciences Institute (SAMSI), Durham, NC 27708;
and Sun is with the  Department of Mathematics,
University of Central Florida,
Orlando, Florida 32816.
Emails:  jzjiang@guet.edu.cn; cheng87@math.duke.edu;   qiyu.sun@ucf.edu. This work is partially supported by the National Natural Science Foundation of China (Grant No. 61761011),
SAMSI under the National Science Foundation (DMS-1638521), and the  National Science Foundation (DMS-1412413).}}

\maketitle

\begin{abstract} 
In this paper, 
we consider  nonsubsampled graph filter banks (NSGFBs) to process data  on a graph 
in a distributed manner. 
   Given an analysis filter bank with small bandwidth, we propose algebraic and optimization methods of constructing
synthesis filter banks such that the corresponding NSGFBs provide a perfect signal reconstruction
in the noiseless setting. Moreover, we prove that the proposed NSGFBs can control the resonance effect in the
presence of bounded noise  and
 they  can limit the influence of shot noise   primarily 
  to a small neighborhood of its location on the graph.
  For an NSGFB on a graph of large size, 
a
distributed implementation has  a significant advantage, since
data processing and communication demands for the agent at
each vertex depend mainly on its neighboring topology.
In this paper, we propose  an iterative distributed algorithm to implement the proposed NSGFBs.
Based on NSGFBs, we also develop a  distributed
denoising technique which is demonstrated to have satisfactory performance on
noise suppression.
   \end{abstract}


\vskip-1.5mm

\section{\bf Introduction}  

  Spatially distributed networks (SDNs)  have an agent at each location equipped with some data processing and communication abilities,
  and they
 have been widely used in  wireless sensor networks, power grids and  many   real world applications
(\cite{aky02}--\cite{cheng_SDS16}).
Data  collected by an SDN
 resides naturally
 on vertices  of a graph. 
Graph signal processing  provides an innovative 
 framework to process  data  on graphs.
Many  concepts, such as the Fourier transform, wavelet transform and filter banks, in classical signal processing,
 have been extended to graph settings in recent years.
  However there are still lots of 
  fundamental
   problems 
    unexplored or
    not completely answered  
(\cite{chungBook2006}--\cite{sandryhaila14}).

\begin{figure}[h]
\begin{center}
\includegraphics[width=50mm, height=15mm]{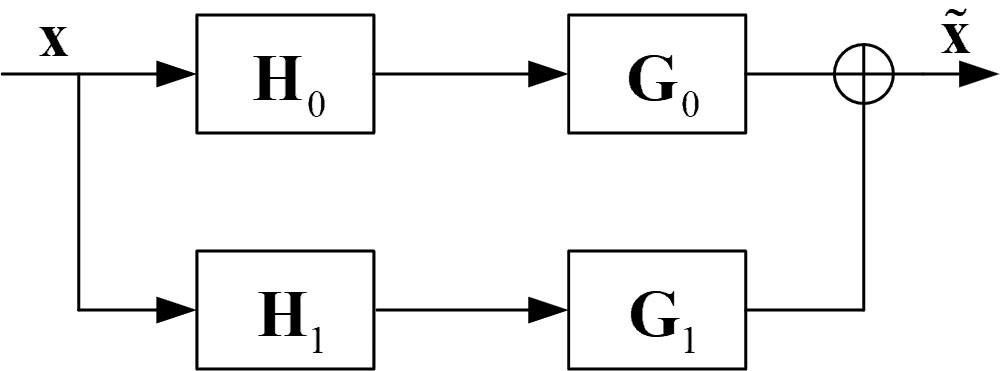}
\caption{Block diagram of an  NSGFB  with analysis filter bank
$({\bf H}_0, {\bf H}_1)$ and  synthesis filter bank $({\bf G}_0, {\bf G}_1)$, where  ${\bf x}$ is the input of the NSGFB and  $\tilde {\bf x}$ is its output.}
\label{filterbank_structure}
\end{center}
\end{figure}

The wavelet transform is one of the most prominent techniques to process signals in  regular domains
 (\cite{daubechiesbook}--\cite{vetterlibook}).
  During the past decades,  graph wavelet transforms have been introduced and
some of them
are
designed using the eigenvalue and eigenspace information of the graph Laplacian matrix
 (\cite{crovella03}--\cite{hammod11}).
 Graph wavelet transform is under the same theoretical structure with  graph filter banks  
 and the  corresponding wavelet filter bank
carries  a down and up-sampling procedure
 (\cite{coifman2008, shuman13}, \cite{narang12}--\cite{tremblay16}).
Several forms of the down and up-sampling procedure have been defined by
the partitioned graph coloring in  \cite {narang12},
the maximum spanning tree structure of the graph in \cite {nguyen15},
and   the SVD decomposition of the graph Laplacian matrix in \cite{shuman13b}.
A proper definition of the down and up-sampling procedure is not obvious
 especially  when the residing graph is of large size and  complicated topological structure.
 This motivates us to consider
    a {\em nonsubsampled graph filter bank} (NSGFB)
that contains an analysis filter bank
$({\bf H}_0, {\bf H}_1)$ and a synthesis filter bank $({\bf G}_0, {\bf G}_1)$, see  Figure \ref{filterbank_structure} for its block diagram.
       The analysis filter bank decomposes a graph signal into two components carrying different frequency information.
          The nonsubsampled structure in an NSGFB greatly simplifies the design of analysis filter banks for spectral decomposition and
   synthesis filter banks for signal reconstruction.

Filter banks  can be implemented  either in a centralized system or a cooperative decentralized (distributed) system. 
In a centralized system, a central facility receives  data from agents at  vertices, performs designed data processing and sends the processed data back to agents at  vertices.  In a decentralized system,  the agent at each vertex has certain data processing ability
 to perform designed data processing, and data  collected from an agent at each vertex is shared only with neighboring vertices.
 Most filter banks on
 graphs  
 are designed 
 for centralized processing, however for the implementation of filter banks on a graph of large size, a centralized system may suffer from high computational burden
 and    call for significant efforts  to create a  data exchange network.
   For signal processing on an SDN or a  graph of large size, 
  a distributed  implementation provides an indispensable tool.
  It has been used for signal sampling and reconstruction on an SDN in  \cite{cheng_SDS16}, graph signal inpainting in \cite{siheng15}
 and  economic dispatch in power networks \cite{doan16}. The reader may refer to  \cite{cheng_SDS16}, \cite{siheng15}--\cite{zhang16} and references therein
  on distributed implementation of signal processing on graphs.  
        In a distributed implementation of the analysis and synthesis procedures of an NSGFB, 
        signal information on each vertex is transmitted only
        to neighboring vertices, which dramatically reduces the computational cost and
calls for low energy consumption.
In this paper, we study   NSGFBs on a cooperative decentralized system  from design to distributed implementation,  and then to distributed signal denoising.

\subsection{Main contributions} 

An important concept for an NSGFB  
is the {\em perfect reconstruction} condition, i.e., the output  $\tilde {\bf x}$ in Figure \ref{filterbank_structure} is always the same as the input  ${\bf x}$, which can be characterized by
 the following matrix equation,
\begin{equation}\label{biorthogonal.condition}
{\bf G}_0{\bf H}_0+{\bf G}_1{\bf H}_1={\bf I},
\end{equation}
 where  $({\bf H}_0, {\bf H}_1)$ and  $({\bf G}_0, {\bf G}_1)$ are its
 analysis   and  synthesis filter bank  respectively.
  Given an analysis filter bank, the existence of synthesis filter banks  is theoretically guaranteed so that the corresponding
   NSGFB satisfies the perfect reconstruction condition \eqref{biorthogonal.condition}  (\cite{mallatbook, vetterlibook}).
    The first contribution of this paper is that we introduce two methods to construct localized synthesis filter banks.
 In the first approach, the synthesis filter bank is obtained by solving a Bezout identity for
polynomials.    Its bandwidth could be no larger
than the bandwidth of the analysis filter bank. 
 In the second approach,
the synthesis filter bank is the solution of a constrained optimization problem.
It 
does not necessarily have small bandwidth, however  it 
 has an exponential off-diagonal decay.
 Consequently,
 the output of the corresponding localized NSGFB suffers  primarily 
 in a small neighborhood of vertices where agents lose data processing  ability and/or communication capability.

    In some real world applications of an NSGFB,
  the input  ${\bf x}$  is  the original signal ${\bf x}_o$ corrupted by an additive noise ${\pmb \epsilon}$. In addition,
  the subband signals  ${\bf z}_0={\bf H}_0 {\bf x}$ and ${\bf z}_1={\bf H}_1 {\bf x}$
   are usually processed via some (non)linear
procedure, such as hard/soft thresholding 
and quantization.
 Then the output
 \begin{equation}\label{noisyoutput.def0}
\tilde {\bf x}= {\bf G}_0 \Psi_0( {\bf H}_0 ({\bf x}_o+\pmb \epsilon))+
{\bf G}_1 \Psi_1( {\bf H}_1 ({\bf x}_o+\pmb \epsilon))
\end{equation}
of the NSGFB is no longer the original signal ${\bf x}_o$, where
${\pmb \epsilon}$ 
is  the  input  noise, 
and
 $\Psi_0, \Psi_1$ are subband processing operators.
 The robustness of
 an NSGFB 
 is of paramount importance.
For an SDN, an agent at each vertex operates almost independently and the
 noise  that arises at each vertex of the graph is usually contained in some range \cite{cheng_SDS16}.
So we may use a bounded determinstic/random noise model for NSGFBs on a distributed system. A reasonable fidelity measure to assess the robustness of an NSGFB is the bounded difference $\|\tilde {\bf x}-{\bf x}_o\|_\infty$, instead of the conventional least squares error  $\|\tilde {\bf x}-{\bf x}_o\|_2$,
between the original signal ${\bf x}_o$ and the output signal $\tilde {\bf x}$ (\cite{cheng_SDS16},  \cite{wangieee09}, \cite{sunaicm14}).
 Here for $1\le p\le \infty$,
 $\ell^p$ is the space of all $p$-summable sequences with norm  $\|\cdot\|_p$.
 The second contribution of this paper is that for the NSGFB with analysis filter bank having small bandwidth and synthesis filter banks obtained from our approaches,
 we establish a quantitative estimate on
 the bounded difference $\|\tilde {\bf x}-{\bf x}_o\|_\infty$, which is independent on the size of the graph ${\mathcal G}$.
This indicates that the proposed NSGFB  can control the resonance effect in the presence of bounded  additive noises.

 For an NSGFB on a graph  of large size,
  a distributed  implementation may provide an indispensable tool.
The third contribution of this paper is that we propose an iterative distributed algorithm
to implement the synthesis procedure of an
NSGFB  rather than finding the synthesis filter bank explicitly.
 The keys behind the algorithm are
   the decomposition \eqref{decomposition}  that splits  the whole residing graph into a family of overlapping subgraphs of appropriate size,  and  an observation that
 solutions of some global optimization problem
 can be locally approximated by  solutions of local optimization problems, when the objective function and  constraints are well localized
 \cite{cheng_SDS16}.
As an application of NSGFBs, we develop a distributed denoising technique that has
satisfactory  performance on noise suppression.


\subsection{Organization}

In Section \ref{graph.section}, we briefly review some fundamental
concepts of  graphs and introduce an overlapping graph decomposition \eqref{decomposition}.
  In Section \ref{filtering.section}, we introduce the concept of  graph filters on $\ell^p, 1\le p\le \infty$, and show that  bounded filters with finite bandwidth
are graph filters on $\ell^p$, see Definition \ref{graphfilterellp.def} and Proposition \ref{finiteband.pr}.
In Section \ref{analysisfilterbanks.section}, we discuss the analysis filter bank $({\bf H}_0, {\bf H}_1)$ of an NSGFB
 which
are required to have small bandwidth, to pass/block the normalized constant signal,  and to have
stability on $\ell^2$, see Assumptions \ref{analysis.assumption1}, \ref{analysis.assumption2} and \ref{analysis.assumption3}.
We show that
analysis filter banks
  have  stability on $\ell^p$ for all $1\le p\le\infty$, with an estimate on their lower and upper $\ell^p$-stability bounds
independent on the size of the graph, see Theorem \ref{filterbankpnorm.thm}.
In Section \ref{perfectreconstruction.section1}, we propose
an algebraic design of  synthesis filter banks  $({\bf G}_0, {\bf G}_1)$
when analysis filters ${\bf H}_0$ and ${\bf H}_1$ are polynomials of the symmetric normalized Laplacian on the graph, see Theorem \ref{bezout.thm}.
In Section \ref{synthesisfilterbank.section2}, we consider
 the construction of   synthesis filter banks  $(\mathbf{G}_{0}, {\bf G}_1)$
 by solving the constrained optimization problem
\eqref{frobenius_optimization} and \eqref{frobenius_optimization2}
 with
the objective function consisting of Frobenius
norms of  ${\bf G}_0$ and ${\bf G}_1$, see \eqref{pseudoinverse} and Theorem \ref{synthesisdecay.thm}.
  In Section  \ref{ida.section}, we propose an exponentially convergent iterative algorithm  \eqref{iterativedistributedalgorithm.eqm}
 and \eqref{iterativedistributedalgorithm.eq0} to implement the synthesis procedure, where each iteration can be implemented in a distributed manner, see Theorem \ref{convergence.thm} and Algorithm \ref{distributed_cheng.algorithm}. %
In Section \ref{simulations.section}, we create a distributed denoising technique associated with  spline NSGFBs, see Figure \ref{denoising_structure}, and
demonstrate
its 
 performance 
for  signal denoising on graphs of large size.
All proofs are collected in the appendices.

\subsection{Notation}

 We use the common convention of representing matrices and vectors with bold letters and scalars with normal letters.
For a matrix ${\bf A}$, denote its transpose, trace, Frobenius norm and operator norm on $\ell^p, 1\le p\le \infty$, by ${\bf A}^T, \operatorname{tr} ({\bf A}), \|{\bf A}\|_F$ and $\|{\bf A}\|_{{\mathcal B}_p}$ respectively.
  For a graph ${\mathcal G}$, denote its adjacency matrix and degree matrix   by ${\bf A}_{\mathcal G}$ and ${\bf D}_{\mathcal G}$
respectively, and define its Laplacian matrix  by
 ${\bf L}_{\mathcal G}:={\bf D}_{\mathcal G} -{\bf A}_{\mathcal G}$
 and its symmetric normalized Laplacian matrix by
 ${\bf L}_{\mathcal G}^{\rm sym}:={\bf D}_{\mathcal G}^{-1/2} {\bf L}_{\mathcal G }{\bf D}_{\mathcal G}^{-1/2}$.
 For a scalar $t$, let ${\rm sgn}(t)$, $\lfloor t\rfloor$ and $t_+$   be its sign, integral part and positive
part respectively,   and ${\bf t}$ be the vector of appropriate size with all entries taking value  $t$. For a set $F$, denote
its cardinality and indicator function   by  $\# F$ and $\chi_F$ respectively.

\section {Preliminaries on graphs}
\label{graph.section}

Let ${\mathcal G}:=(V,E)$ be a graph,
where $V =\{1, 2, \cdots, N\}$
is the set of vertices 
 and $E$ is the set of edges  (\cite{cheng_SDS16, chungBook2006}). 
For the distributed implementation of an  NSGFB, we  require that the residing graph  ${\mathcal G}$ has certain global  features.

 \begin{assumption}\label{assumption1}  \rm   Throughout the paper, we  consider simple graphs ${\mathcal G}$, i.e., they are undirected and  unweighted, and
 they do not contain self-loops  and  multiple edges.
\end{assumption}

Take $r\ge 0$.
For a graph ${\mathcal G}=(V,E)$ satisfying Assumption \ref{assumption1},
 we define the  {\em $r$-neighborhood} and
 the
   {\em $r$-neighboring subgraph} of  $i\in V$ by
 $B(i,r):=\big\{j \in V: \ \rho(i,j) \leq r \big\}$
  and
${\mathcal G}_{i,  r}:= \big( B (i, r),  E(i, r)\big)$
respectively, where
$E(i,r)$ contains all edges of the graph ${\mathcal G}$
with endpoints in $B(i,r)$, and $\rho(i,j)$ is the geodesic distance between vertices $i$ and $j$ in $V$.
Then for $r\ge 1$,  we can decompose the graph   ${\mathcal G}$ 
into  a family of overlapping
subgraphs ${\mathcal G}_{i,  r},  i\in V$, of diameters at most $2r$,
\begin{equation}\label{decomposition}
{\mathcal G}=\cup_{i\in V}  {\mathcal G}_{i,  r}.\end{equation}
For  distributed implementation for  an  NSGFB, 
  we  presume that numbers of vertices in the
$r$-neighborhood of any vertex  are
dominated by a polynomial about $r$.

   \begin{assumption}\label{assumption2} \rm  Throughout the paper, we  consider graphs
   ${\mathcal G}$ with the counting measure  $\mu$
    having polynomial growth, i.e.,
   there exist positive constants $D_1({\mathcal G})$ and $d$ such that
   \begin{equation}\label{countmeasure.pr.eq1}
   \mu(B(i,r))\le D_1({\mathcal G})(r+1)^{d} \end{equation}
   for all $i\in V$ and $r\ge 0$,
     where $\mu(F):=\# F$ for all $F\subset V$.
\end{assumption}

The minimal constants $d$ and $D_1({\cal G})$ in \eqref{countmeasure.pr.eq1} are called  as {\it Beurling dimension} and {\it  density} of the graph ${\mathcal G}$ respectively \cite{cheng_SDS16}.

\begin{figure}[h]  
\begin{center}
\includegraphics[width=42mm, height=38mm]{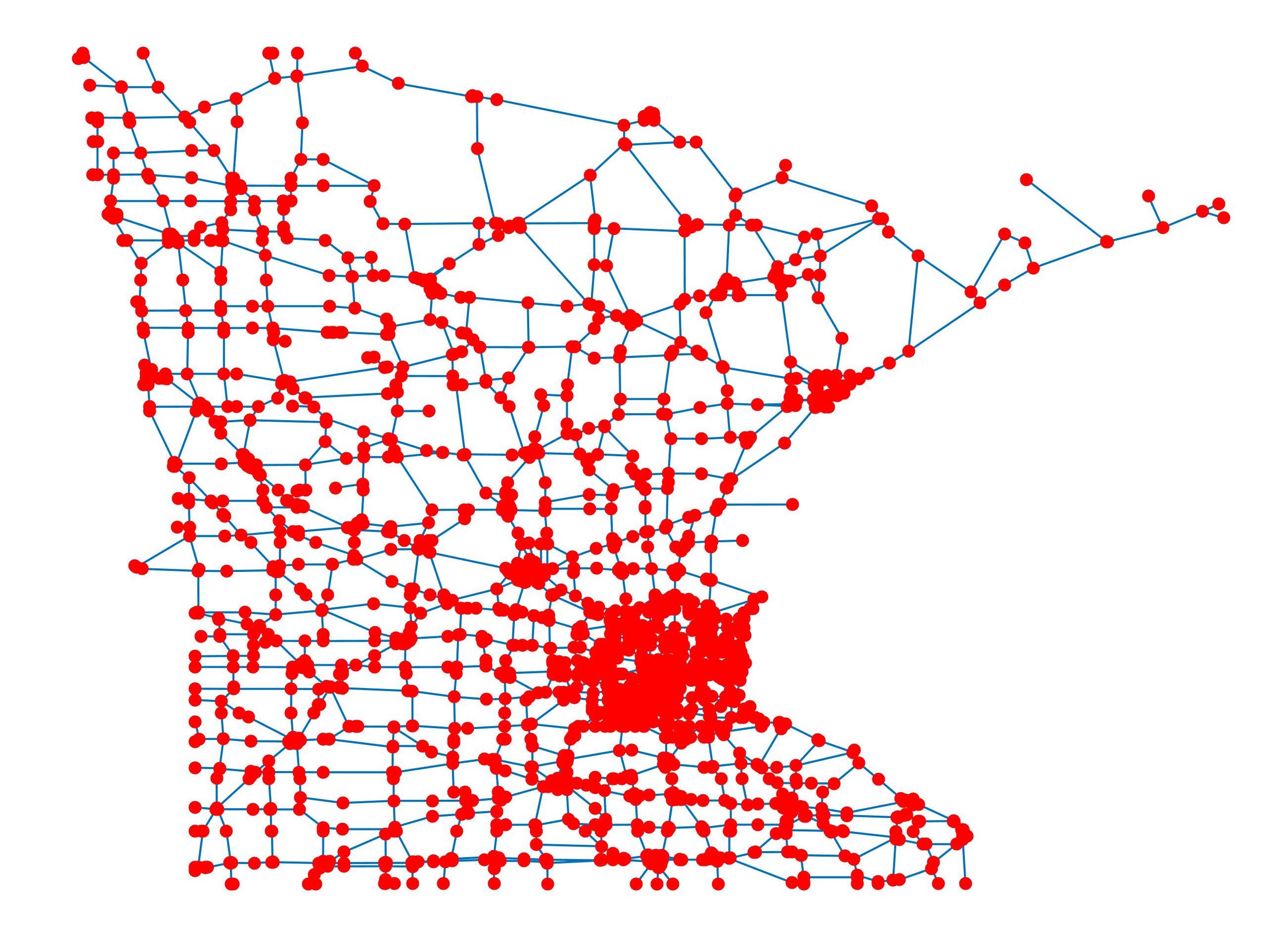}
\includegraphics[width=42mm, height=38mm]{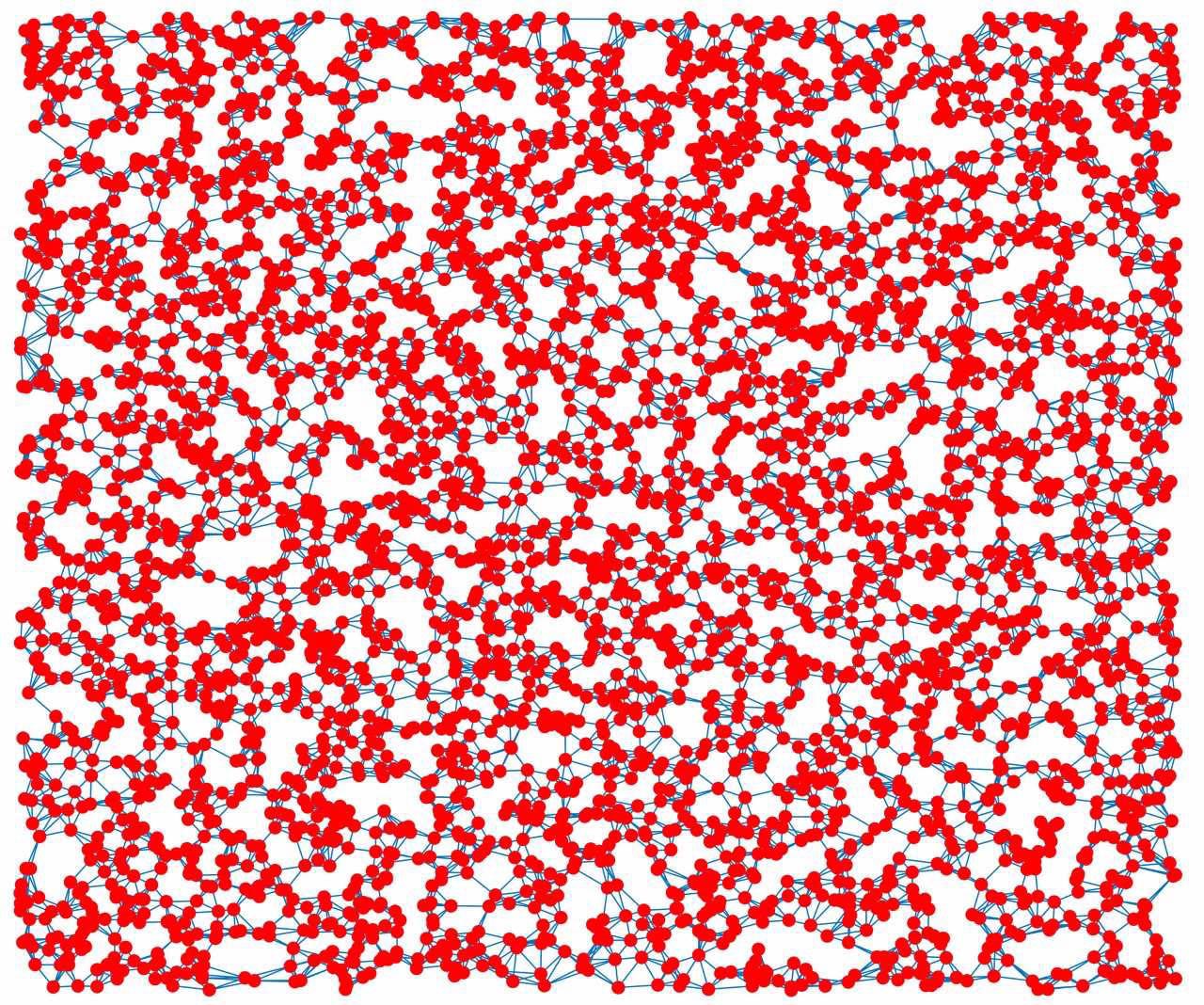}
\caption{Plotted 
on the left  is  the Minnesota  traffic graph that has Beurling dimension $2$ and
density $2.1378$. On the right is a random geometric graph 
 with $N=4096$, which has
Beurling dimension $2$ and
density $3.0775$. }
\label{circulant_graph}
\end{center}
\end{figure}

\smallskip
  The decomposition  \eqref{decomposition}  plays a crucial role in
the proposed distributed implementation for an NSGFB,
and
the selection of the  radius parameter $r$ in \eqref{decomposition}
  depends on   Beurling dimension $d$ and density $D_1(\cal G)$ of the graph ${\mathcal G}$, see Theorem \ref{convergence.thm}.
 Accordingly, we expect that the Beurling dimension $d$ and density $D_1(\cal G)$ of the graph ${\mathcal G}$ are much smaller than (or even independent on) the size  of the graph,
 which implies that the graph ${\mathcal G}$ should be sparse.
 Shown in Figure \ref{circulant_graph} are two representative
graphs that satisfy Assumptions \ref{assumption1}  and \ref{assumption2}: 
\begin{itemize}
\item The Minnesota traffic graph with 2642 vertices, where each vertex represents a spatial location in the state of Minnesota
equipped with a traffic monitoring sensor and each edge denotes a direct
communication link between monitoring sensors 
(\cite{narang12, narang13}).

\item
The random geometric graph ${\rm RGG}_N$ with $N$
vertices  randomly deployed in the region $[0, 1]^2$ and an edge  existing
between two vertices if their physical distance is not larger
than $\sqrt{2} N^{-1/2}$   (\cite{shuman13b, shuman16, penrose2003}). 
\end{itemize}

\section {Graph signal and filtering}
\label{filtering.section}

Let $ {\mathcal G}=(V,E) $ satisfy Assumptions \ref{assumption1} and \ref{assumption2}.
A {\em  signal} ${\bf x}$  residing on the graph ${\mathcal G}$ is a vector $(x_i)_{i\in V}$,  where $x_i$ refers to the signal value at vertex $i\in V$.  In  SDNs and many real world applications,
 data collected belongs to some sequence space  $\ell^p, 1\le p\le \infty$
  (\cite{cheng_SDS16}, \cite{wangieee09}, \cite{sunaicm14}).

A {\em filter} ${\bf A}$ on the graph ${\mathcal G}$  is a linear transformation from one signal ${\bf x}$ on ${\mathcal G}$
 to another signal  ${\bf y}={\bf A} {\bf x}$ on ${\mathcal G}$, which is usually represented by a  matrix
${\bf A}=(a(i,j))_{i,j\in V}$.  A filter  ${\bf A}$ 
is expected to map a signal with finite energy to another signal with finite energy and
a bounded signal to another bounded signal.  A quantitative description of the above filtering procedure is
\begin{equation}\label{filter.def0}
 \|{\bf A}{\bf x}\|_p  \leq C \|{\bf x}\|_p \ {\rm for \ all}\  \ {\bf x}\in \ell^p,
\end{equation}
where $1\le p\le \infty$ and $C$ is a positive constant.

\begin{definition}\label{graphfilterellp.def}\rm  Let $1\le p\le \infty$. We say that  ${\bf A}$  is a   {\em  graph filter on $\ell^p$}
 if \eqref{filter.def0} is satisfied, and we call the minimal constant  $C$ for \eqref{filter.def0} to hold, denoted by $\|{\bf A}\|_{{\mathcal B}_p}$,
 the  {\em filter bound}  on $\ell^p$.
\end{definition}


 In some practical applications  (\cite{shuman13, narang12, narang13, tanaka14,  shuman13b,  dragotti2017}), a graph filter ${\bf A}$ is a polynomial  $P(t):=\sum_{l=0}^L p_l t^l$ of
the symmetric normalized Laplacian   ${\bf L}_{\mathcal G}^{\rm sym}$ on ${\mathcal G}$, i.e.,
\begin{equation} \label{polynomialfilter.def}
{\bf A}=P({\bf L}_{\mathcal G}^{\rm sym})=p_0{\bf I}+ \sum_{l=1}^L p_l ({\bf L}_{\mathcal G}^{\rm sym})^l.
\end{equation}
Let $0\le \lambda_1\le \lambda_2\le\cdots\le \lambda_N\le 2$  be eigenvalues of the symmetric normalized Laplacian   ${\bf L}_{\mathcal G}^{\rm sym}$
and write
 \begin{equation}\label{laplacian.decomposition}
 {\bf L}_{\mathcal G}^{\rm sym}=  {\bf U}^T {\pmb \Lambda} {\bf U},\end{equation}
  where ${\bf U}^T=[{\bf u}_1, \ldots, {\bf u}_N]$ is an orthogonal matrix and $\pmb \Lambda={\rm diag} (\lambda_1, \ldots, \lambda_N)$
 is a diagonal matrix.
 Then
 \begin{equation} \label{polynomialfilter.eigenvalue}
{\bf A}= {\bf U}^T P({\pmb \Lambda}) {\bf U},
\end{equation}
and  the  filter bound $\|{\bf A}\|_{{\mathcal B}_2}$  
can be evaluated explicitly, 
\begin{equation}\label{polynomialfilter.eig}
\|{\bf A}\|_{{\mathcal B}_2}=\sup_{1\le n\le N} |P(\lambda_n)|\le \sup_{0\le t\le 2} |P(t)|.
\end{equation}

To estimate $\|{\bf A}\|_{{\mathcal B}_p}, p\ne 2$, of  a graph filter ${\bf A}=(a(i,j))_{i,j\in V}$, we define  the  {\em bound} of ${\bf A}$  by
\begin{equation}\label{boundentries.def}
\|{\bf A}\|_{\infty}=\sup_{i,j\in V} |a(i,j)|. 
\end{equation}
A graph filter ${\bf A}$ on $\ell^p, 1\le p\le \infty$,  has bounded entries and
\begin{equation}\label{infty.estimate}
\|{\bf A}\|_\infty 
\le
\sup_{j\in V} \|{\bf A} {\bf e}_j\|_p \le \|{\bf A}\|_{{\mathcal B}_p}, \ 1\le p\le \infty,
\end{equation}
where the last inequality is obtained from \eqref{filter.def0}
by replacing ${\bf x}$  by the standard unit vector  ${\bf e}_j$  with $j$-th component taking value one while all others are zero.

 For the distributed implementation for an  NSGFB,  bounded filters with finite bandwidth will be used as analysis filters, see Assumption \ref{analysis.assumption1}.

\begin{definition}\label{bandwidth.def}  \rm
The {\em bandwidth}  $\sigma:=\sigma({\bf A})$ of a graph filter ${\bf A}=(a(i,j))_{i,j\in V}$  is the minimal nonnegative integer such that $a(i,j)=0$ for all $i,j\in V$ with $\rho(i,j)>\sigma$.  For a  
filter pair $({\bf A}, {\bf B})$, we define its bandwidth  $\sigma:=\sigma({\bf A}, {\bf B})$ by $\max(\sigma({\bf A}), \sigma({\bf B}))$.
\end{definition}

In the following proposition, we show that a bounded filter with finite bandwidth  is a  graph filter on $\ell^p, 1\le p\le \infty$,
with  filter bound dominated by some constant, independent of the size of the graph ${\mathcal G}$.

\begin{proposition}\label{finiteband.pr}  \rm  Let $1\le p\le \infty$,  and ${\mathcal G}$ be a  graph satisfying Assumptions \ref{assumption1} and \ref{assumption2}.  Then for any  bounded graph filter  ${\bf A}$  with bandwidth $\sigma$, we have
\begin{equation}\label{finiteband.pr.eq1}
\|{\bf A}\|_{\infty}\le \|{\bf A}\|_{{\mathcal B}_p}
\le D_1({\mathcal  G})  (\sigma+1)^d  \|{\bf A}\|_{\infty},
\end{equation}
where $d$ and $D_1({\mathcal G})$ are the
Beurling dimension and density of the graph ${\mathcal G}$ respectively.
\end{proposition}

For $n\ge 1$,  we define
 {\em spline  filters  ${\bf H}_{0, n}^{\rm spln}$ and ${\bf H}_{1, n}^{\rm spln}$ of order $n$} by
\begin{equation} \label{lowpassspline.def}
\mathbf{H}_{0, n}^{\rm spln}= \Big({\bf I}-\frac{1}{2} {\bf L}_{\mathcal G}^{\rm sym}\Big)^n
 \ \ {\rm and }\  \ \mathbf{H}_{1, n}^{\rm spln}=\Big(\frac{1}{2}{\bf L}_{\mathcal G}^{\rm sym}\Big)^n,
\end{equation}
 see \cite{dragotti2017}
in circulant graph setting.
 Spline  filters  ${\bf H}_{0, n}^{\rm spln}$ and ${\bf H}_{1, n}^{\rm spln}, n\ge 1$,
have bandwidth $n$
and their filter bounds on $\ell^2$ dominated by one, i.e.,
\begin{equation}\label{spline.ell2norm}
\|{\bf H}_{l, n}^{\rm spln}\|_{{\mathcal B}_2}\le 1, \ l=0, 1. \end{equation}
For $p\ne 2$,  we obtain from Proposition \ref{finiteband.pr}  that
\begin{eqnarray}
\|{\bf H}_{l, n}^{\rm spln}\|_{{\mathcal B}_p} & \hskip-0.08in \le & \hskip-0.08in  \|{\bf H}_{l, 1}^{\rm spln}\|_{{\mathcal B}_p}^n\le
\big(2^{d}D_1({\mathcal  G}) \|{\bf H}_{l, 1}^{\rm spln}\|_\infty )^n\nonumber\\
& \hskip-0.08in = & \hskip-0.08in
\big(2^{d-1}D_1({\mathcal  G}))^n, \ l=0, 1,
\end{eqnarray}
where $d$ and $D_1({\mathcal G})$ are the
Beurling dimension and density of the graph ${\mathcal G}$ respectively.
Therefore 
our representative
spline filters  ${\bf H}_{0,n}^{\rm spln}$ and ${\bf H}_{1, n}^{\rm spln}, n\ge 1$,  are graph filters on $\ell^p, 1\le p\le \infty$, with filter bounds dominated by some constants
 independent of the size   of the graph ${\mathcal G}$.

\section{Analysis filter banks}
\label{analysisfilterbanks.section}

The analysis filter bank decomposes the input signal on a graph into two components carrying 
 frequency information.
In this section, we  design the analysis filter bank $({\bf H}_0, {\bf H}_1)$ of an NSGFB
to have small bandwidth, to pass/block the normalized constant signal,  and to have
stability on $\ell^2$, see Assumptions \ref{analysis.assumption1}, \ref{analysis.assumption2} and \ref{analysis.assumption3}.
In this section, we also show that
analysis filter banks
  have  stability on $\ell^p$ for all $1\le p\le\infty$, with an estimate on their lower and upper $\ell^p$-stability bounds
independent of the size of the graph, see Theorem \ref{filterbankpnorm.thm}.

Let ${\mathcal G}=(V, E)$  be a graph satisfying Assumptions \ref{assumption1} and \ref{assumption2},
and  $({\bf H}_0, {\bf H}_1)$ be the analysis filter bank
 of an NSGFB.
For the distributed implementation of an NSGFB,  we make the following assumption 
for its analysis filter bank $({\bf H}_0, {\bf H}_1)$.

\begin{assumption}\label{analysis.assumption1}  \rm
The analysis filter bank $({\bf H}_0, {\bf H}_1)$  has bandwidth $\sigma\ge 1$.
\end{assumption}

Given an input  graph signal ${\bf x}=(x_i)_{i\in V}$,  outputs of
  analysis procedure are
  \begin{equation}\label{analysisoutput}
{\bf z}_0={\bf H}_0 {\bf x} \ \ {\rm and}\ \ {\bf z}_1={\bf H}_1{\bf x}.\end{equation}
Write ${\bf z}_l=(z_l(i))_{i\in V}$ and ${\bf H}_l=(h_l(i,j))_{i,j\in V}, l=0, 1$. Then it follows from
\eqref{analysisoutput} and Assumption \ref{analysis.assumption1} that component
 values of the outputs  ${\bf z}_0$ and ${\bf z}_1$
  at each vertex $k\in V$
  are weighted sums of values of the input  ${\bf x}$ in a $\sigma$-neighborhood of $k$,
\begin{equation}\label{analysis.implementation}
z_l(k)=\sum_{\rho(i,k)\le \sigma} h_l(k,i) x(i), \ \  i\in V.
\end{equation}
Thus  the analysis procedure of an NSGFB can be implemented in a distributed manner.

\smallskip

To apply  an NSGFB to some real world applications, such as
 noise suppression and abnormal phenomenon detection, 
  its  analysis filter
bank 
should constitute  certain spectral decomposition
(\cite{narang12, narang13, venkatesan15, sakiyama14}).
 Throughout the paper, we also make the following assumption for the analysis filter bank $({\bf H}_0, {\bf H}_1)$.

\begin{assumption}\label{analysis.assumption2}  \rm
The  filter ${\bf H}_0 $
  passes  the  normalized constant signal $ {\bf D}_{\mathcal G}^{1/2}{\bf 1}$, and the filter
  ${\bf H}_1$ blocks  the  normalized constant signal $ {\bf D}_{\mathcal G}^{1/2}{\bf 1}$, i.e.,
  \begin{equation}
  \label{analysis.assumption2.eq}
  {\bf H}_0  {\bf D}_{\mathcal G}^{1/2} {\bf 1}= {\bf D}_{\mathcal G}^{1/2}{\bf 1}\ \ {\rm  and}\  \ {\bf H}_1 {\bf D}_{\mathcal G}^{1/2} {\bf 1}={\bf 0}.\end{equation}
\end{assumption}

The frequency partition of an analysis filter bank on  an arbitrary graph ${\mathcal G}$  is not as obvious 
  as that in  classical setting.
For the case that 
\begin{equation}\label{pop1.filter}
{\bf H}_0= P_0({\bf L}_{\mathcal G}^{\rm sym})\ \ {\rm and}\  \ {\bf H}_1= P_1({\bf L}_{\mathcal G}^{\rm sym})
\end{equation}
for some polynomials $P_0$ and $P_1$, one may verify that 
 Assumption \ref{analysis.assumption2} is satisfied
if and only if
\begin{equation}\label{p0p1.eq} P_0(0)=1\ \ {\rm and }\  \ P_1(0)=0.
\end{equation}
The above equivalence follows from the fact that   ${\bf D}_{\mathcal G}^{1/2}{\bf 1}$
is an eigenvector of  the symmetric normalized Laplacian ${\bf L}_{\mathcal G}^{\rm sym}$   associated with eigenvalue zero.

\begin{figure}[h]  
\begin{center}
\includegraphics[width=28mm, height=28mm]{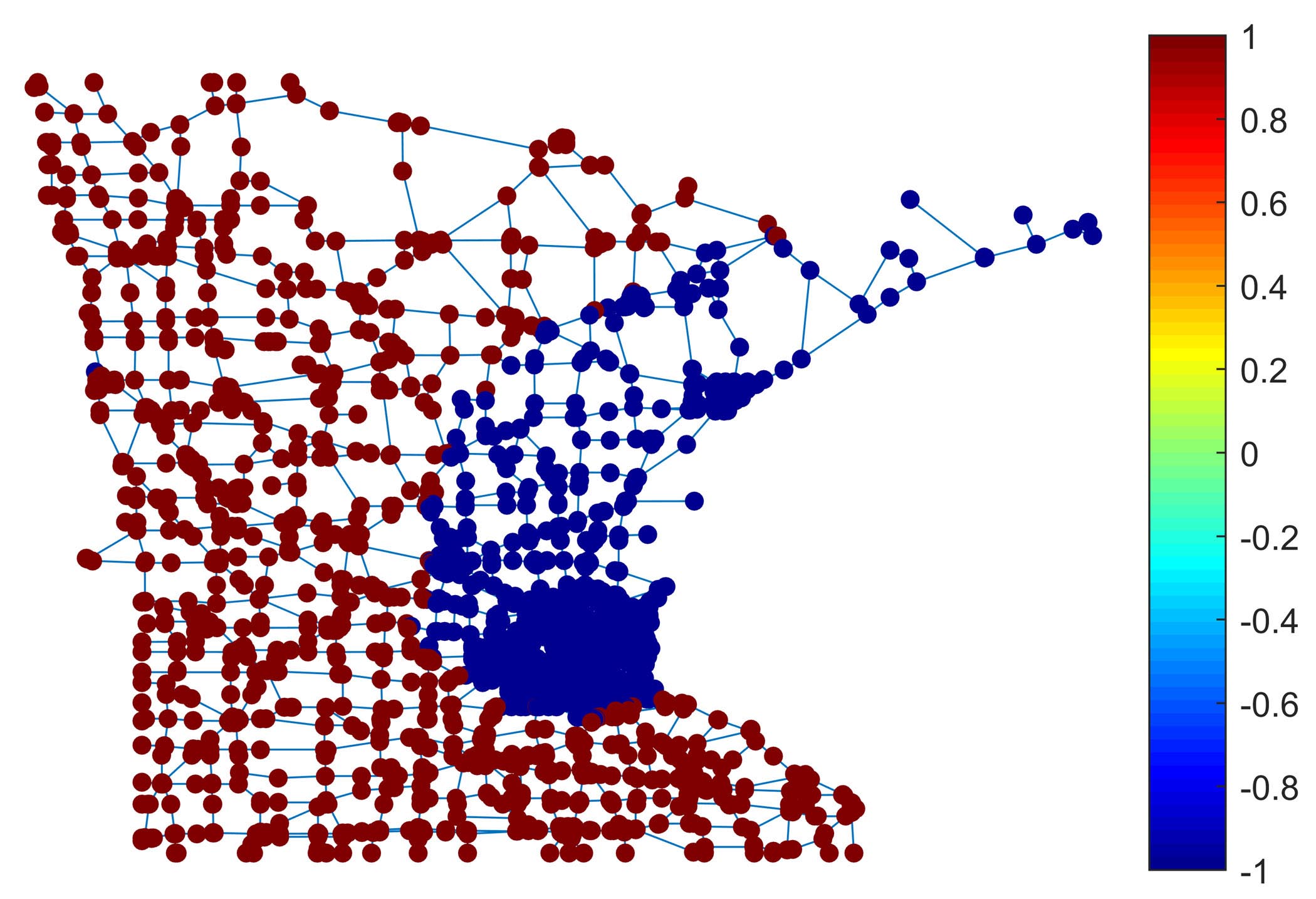}
\includegraphics[width=28mm, height=28mm]{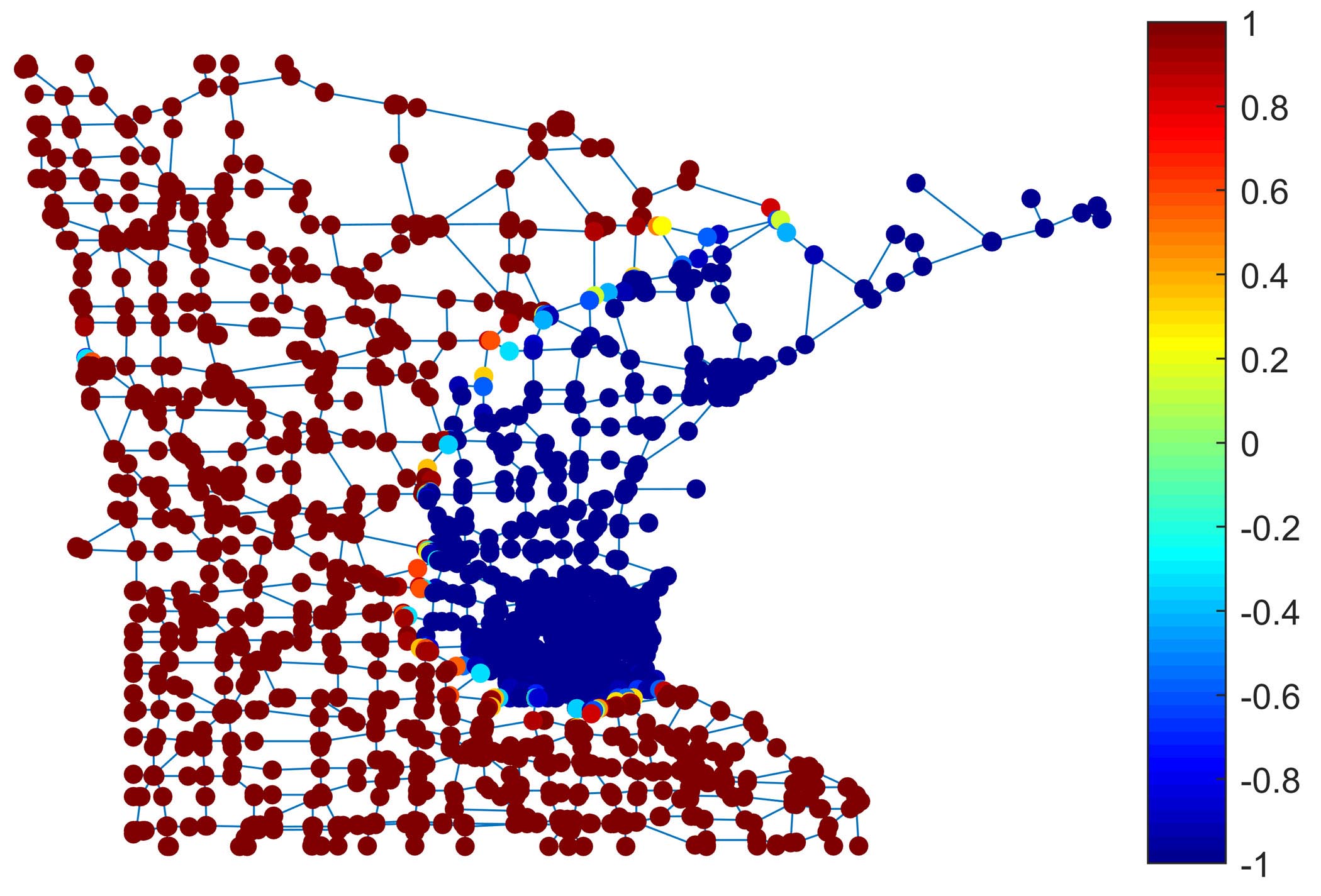}
\includegraphics[width=28mm, height=28mm]{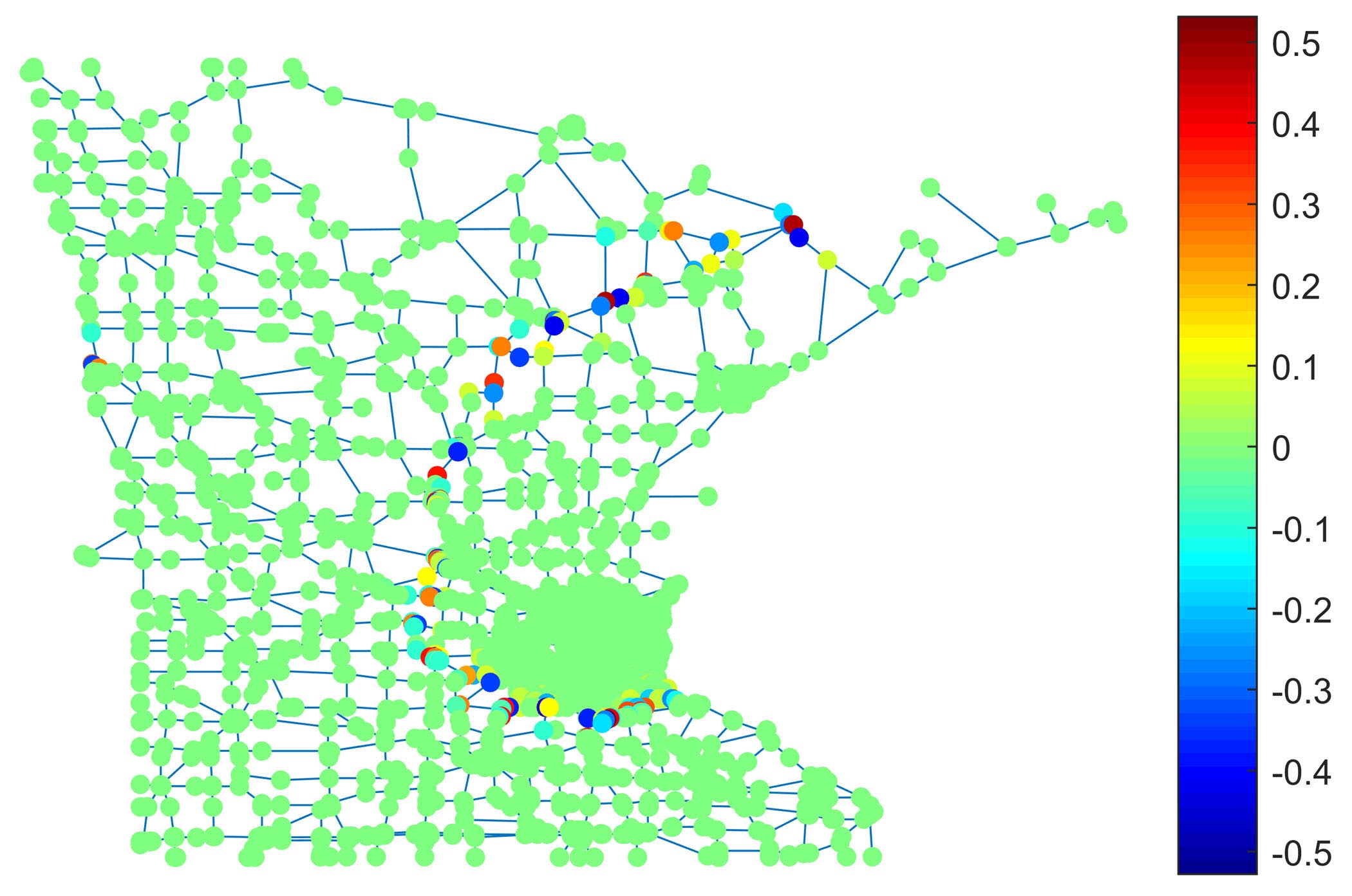}\\
\includegraphics[width=28mm, height=28mm]{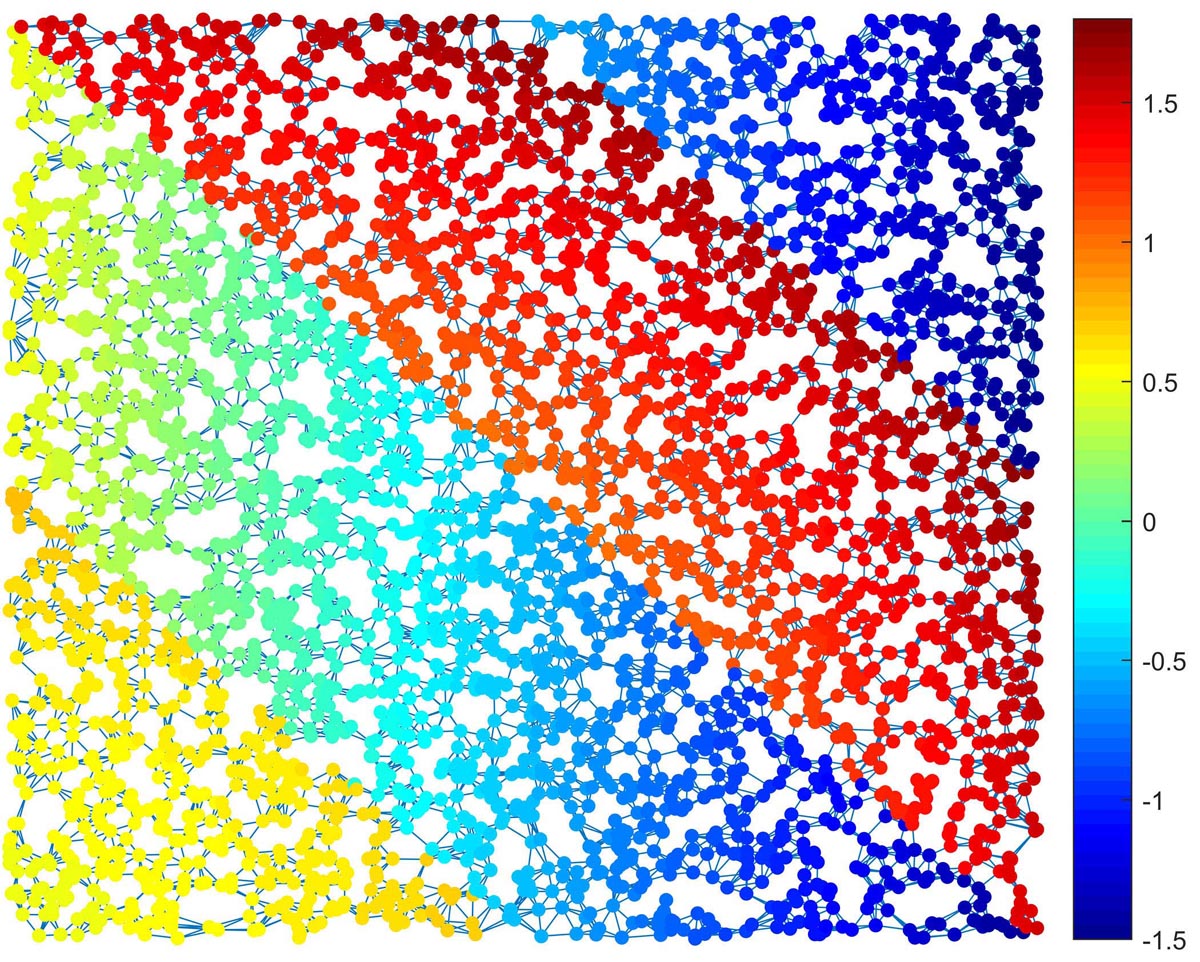}
\includegraphics[width=28mm, height=28mm]{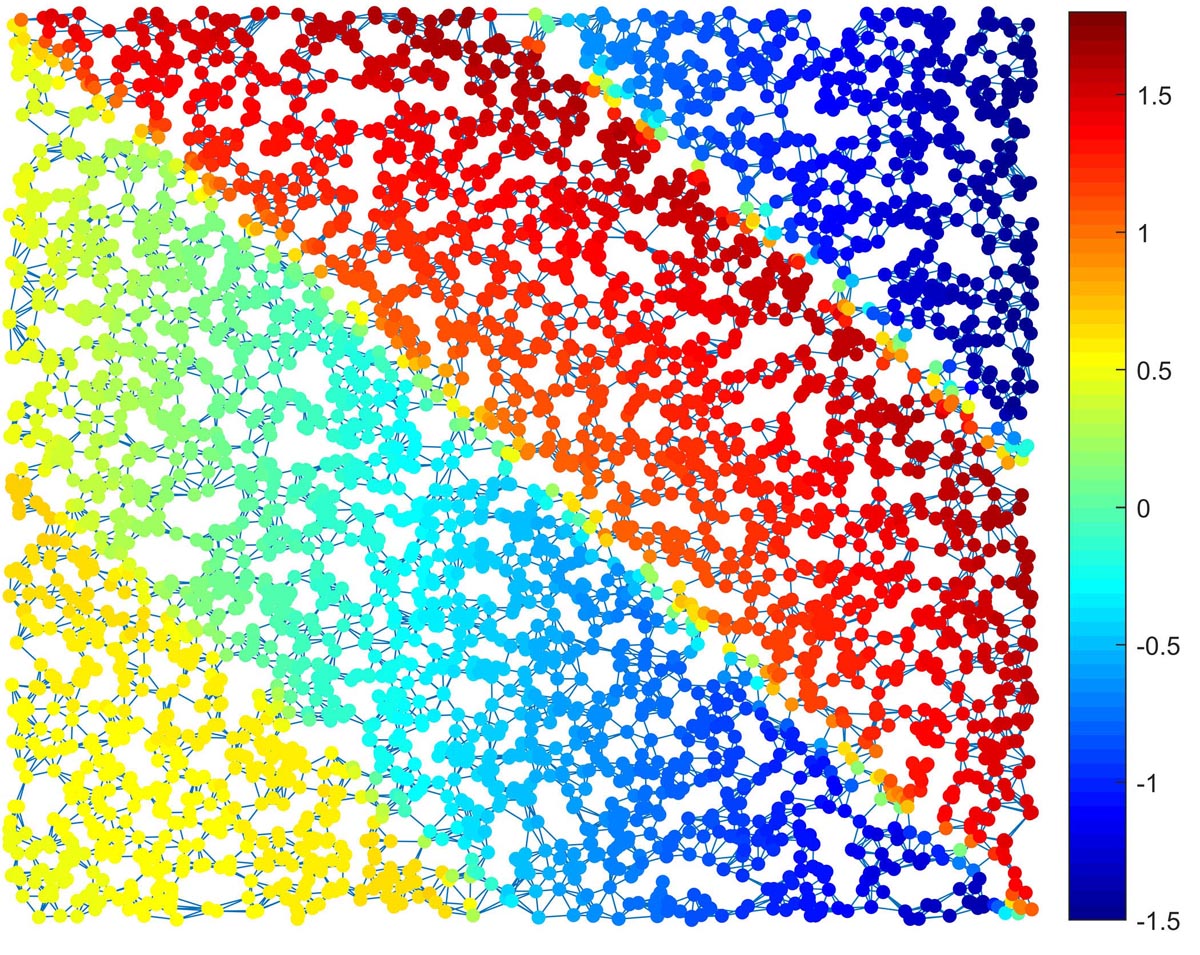}
\includegraphics[width=28mm, height=28mm]{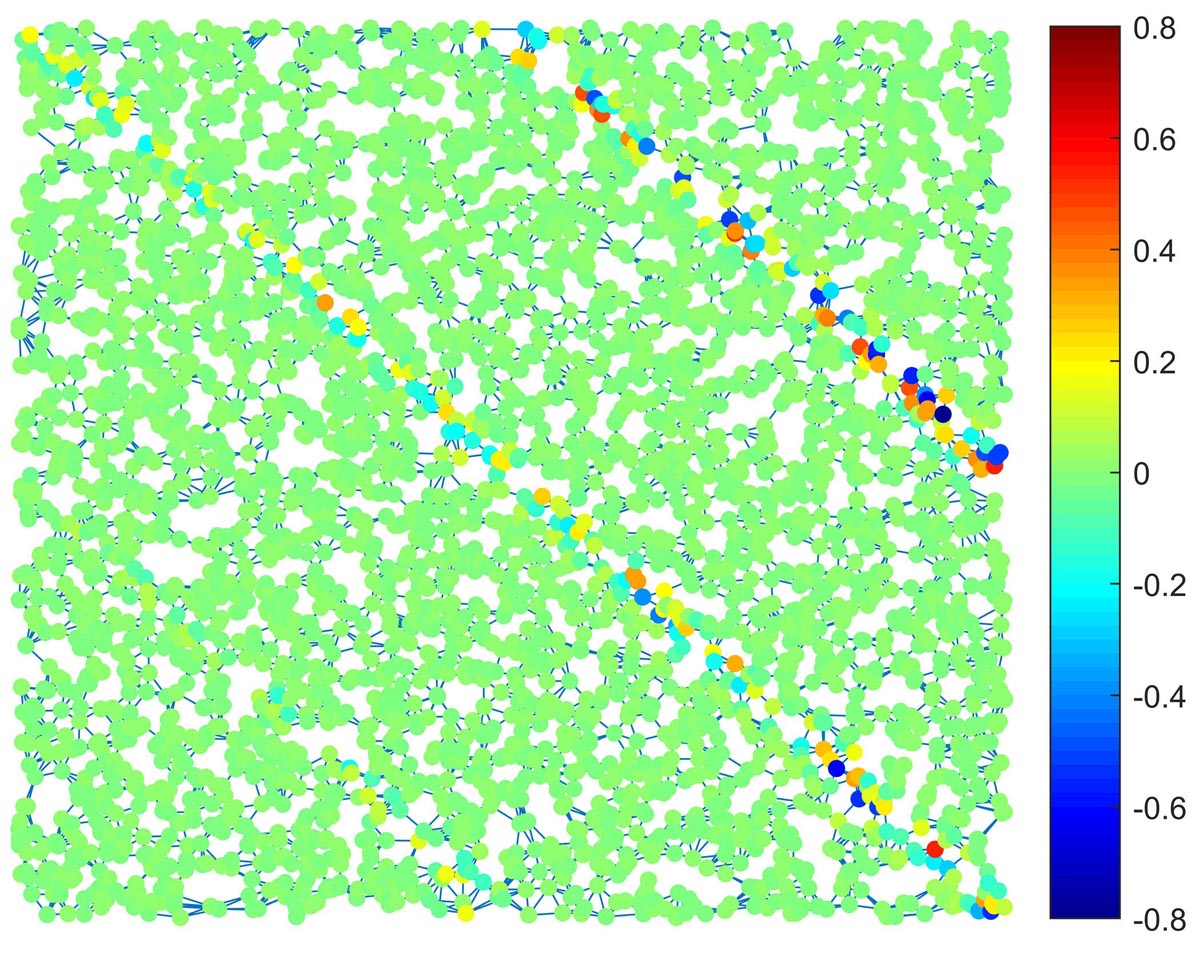}\\
\caption{Plotted on the top (resp. at the bottom),  from  left to  right,  are the original signal ${\bf x}$ on the Minnesota traffic graph
(resp. on the random geometric graph ${\rm RGG}_{4096}$ in Figure  \ref{circulant_graph}),
the lowpass filtered signal ${\bf H}_{0, 2}^{\rm spln} {\bf x}$   and the highpass filtered signal ${\bf H}_{1, 2}^{\rm spln} {\bf x}$.
The signal  ${\bf x}$ on the top is  a blockwise constant function that
 has only two
values  $\pm 1$  on three blocks with one block only containing a vertex (\cite{narang12, narang13}),
and  the  signal ${\bf x}$ at  the bottom  is a blockwise polynomial
 consisting of four strips and imposing the polynomial $0.5-2c_x$ on the first and third diagonal strips and   $0.5+c^2_x+c^2_y$ on the second and fourth strips respectively, where $(c_x,c_y)$ are the coordinates of vertices (\cite{shuman13b}).    }
\label{subband_circulant.fig}
\end{center}
\end{figure}

 The spline filter banks $({\bf H}_{0, n}^{\rm spln}, {\bf H}_{1, n}^{\rm spln}), n\ge 1$, 
   are  of the form \eqref{pop1.filter} with $P_0(t)=(1-t/2)^n$ and $P_1(t)=(t/2)^n$, and
  they satisfy  Assumption \ref{analysis.assumption2} by \eqref{p0p1.eq}, i.e.,
\begin{equation}\label{splinefilterbank.thm.pf.eq0}
 {\bf H}_{0, n}^{\rm spln} {\bf D}_{\mathcal G}^{1/2} {\bf 1}=  {\bf D}_{\mathcal G}^{1/2} {\bf 1}
 \ \ {\rm and}\  \  {\bf H}_{1, n}^{\rm spln} {\bf D}_{\mathcal G}^{1/2} {\bf 1} ={\bf 0}.
 \end{equation}
 Spline filter banks  in the circulant graph setting are known in  \cite{dragotti2017} as graph-spline wavelet transform.
 Shown in  Figure \ref{subband_circulant.fig}  is local smoothing/blocking phenomenon of the spline filter bank $({\bf H}_{0, 2}^{\rm spln}, {\bf H}_{1, 2}^{\rm spln})$  to a blockwise
constant signal on the Minnesota traffic graph and a blockwise smooth signal on the random geometric graph  ${\rm RGG}_{4096}$
in Figure  \ref{circulant_graph}.  It is observed that the lowpass filtered signal is very  close to the original signal except near the boundary between different blocks,
   and that the highpass filtered 
   signal   essentially  vanishes except around the region where the original signal exhibits sharp local variation.

\smallskip

 Robustness is a fundamental requirement in the context of filter bank to control the signal dynamic range and to regulate the input noise.  For the robustness of an NSGFB on $\ell^p, 1\le p\le \infty$, we
 introduce  stability of a graph filter pair  on $\ell^p$.

  \begin{definition}\label{filterbank.def}\rm  Let $1\le p\le \infty$.
  We say that
$({\bf H}_0, {\bf H}_1)$ is {\em stable on $\ell^p$} if
 there are two positive constants $C_p$ and $D_p$ such that
  \begin{equation}\label{filterbank.eq1}
  C_p \| {\bf x}\|_p \leq \big( \|{\bf H}_0 {\bf x}\|^{p}_p+\|{\bf H}_1 {\bf x}\|^{p}_p\big)^{1/p} \leq D_p\| {\bf x}\|_p \end{equation}
  hold for all ${\bf x}\in \ell^p$
  if  $1\le p<\infty$, and
    \begin{equation}\label{filterbank.eq1+} C_\infty \| {\bf x}\|_\infty \leq \max( \|{\bf H}_0 {\bf x}\|_\infty, \|{\bf H}_1 {\bf x}\|_\infty\big) \leq D_\infty\| {\bf x}\|_\infty\end{equation}
hold for all ${\bf x}\in \ell^\infty$ if $p=\infty$.
The optimal constants $C_p$ and $D_p$ for the inequalities in  \eqref{filterbank.eq1} and \eqref{filterbank.eq1+} to hold are called as {\em  lower and upper  stability bounds} of the  graph filter bank $({\bf H}_0, {\bf H}_1)$ on $\ell^p$ respectively.
\end{definition}

Given an NSGFB  with the analysis filter bank $({\bf H}_0, {\bf H}_1)$ and synthesis filter bank
$({\bf G}_0, {\bf G}_1)$ such that the perfect reconstruction condition
\eqref{biorthogonal.condition} holds,  we have that
\begin{eqnarray*}
   (\|{\bf G}_0\|_{{\mathcal B}_2}^2+ \|{\bf G}_1 \|_{{\mathcal B}_2}^2)^{-1}  \|{\bf x}\|_2^2
 & \hskip-0.09in \leq & \hskip-0.09in  \|{\bf H}_0 {\bf x}\|^{2}_2+\|{\bf H}_1 {\bf x}\|^{2}_2 \\
 & \hskip-0.09in \leq & \hskip-0.09in
(\|{\bf H}_0\|_{{\mathcal B}_2}^2+ \|{\bf H}_1 \|_{{\mathcal B}_2}^2)\| {\bf x}\|_2^2
\end{eqnarray*}
hold for all ${\bf x}\in \ell^2$. So throughout the paper, we assume that
the  analysis procedure is stable on $\ell^2$. 

\begin{assumption}\label{analysis.assumption3}  \rm  The analysis filter bank   $({\bf H}_0, {\bf H}_1)$
is  stable on $\ell^2$.
\end{assumption}

For any ${\bf x}\in \ell^2$, direct calculation leads to
\begin{equation}\label{hohl.ell2}
\|{\bf H}_0 {\bf x}\|^{2}_2+\|{\bf H}_1 {\bf x}\|^{2}_2= {\bf x}^T ({\bf H}_0^T {\bf H}_0+{\bf H}_1^T{\bf H}_1\big) {\bf x}.\end{equation}
Thus we have the following characterization to Assumption \ref{analysis.assumption3}.  

\begin{proposition} \label{filterbank.prop}  \rm
Let  ${\mathcal G}$ 
satisfy
 Assumptions \ref{assumption1} and \ref{assumption2}.
Then 
$({\bf H}_0, {\bf H}_1)$  satisfies Assumption \ref{analysis.assumption3} if and only if
  ${\bf H}:={\bf H}_0^T {\bf H}_0+{\bf H}_1^T{\bf H}_1$ is  positive definite.
Moreover,
  the
   optimal  constants  $C_2$ and  $D_2$  for 
   \eqref{filterbank.eq1} to hold
can be evaluated by  \begin{equation}
  C_2^2= (\|{\bf H}^{-1}\|_{{\mathcal B}_2} )^{-1}\ \ {\rm and} \ \ D_2^2 = \|{\bf H}\|_{{\mathcal B}_2}.
  \end{equation}
   \end{proposition}

For graph filters ${\bf H}_0$ and ${\bf H}_1$ of the form
\eqref{pop1.filter}, we obtain from \eqref{laplacian.decomposition} that
 \begin{equation} \label{polynomial.ell2*}
  {\bf H}_{0}^T {\bf H}_{0} + {\bf H}_{1}^T  {\bf H}_{1}= {\bf U}^T \big( (P_0(\pmb \Lambda))^2+(P_1(\pmb \Lambda))^2\big){\bf U}.\end{equation}
Hence    we can evaluate the optimal constants $C_2$ and $D_2$
for 
 (IV.7) to hold explicitly:
  \begin{eqnarray}\label{splinefilterbank.thm.pf.eq3}
 & \hskip-0.08in    & \hskip-0.08in  \inf_{1\le m\le N} (P_0(\lambda_m))^2+(P_1(\lambda_m))^2 \nonumber\\
  & \hskip-0.08in =   & \hskip-0.08in
\inf_{\|{\bf x}\|_2=1} \|{\bf H}_{0} {\bf x}\|^{2}_2+\|{\bf H}_{1} {\bf x}\|^{2}_2
\le \sup_{\|{\bf x}\|_2=1} \|{\bf H}_{0} {\bf x}\|^{2}_2+\|{\bf H}_{1} {\bf x}\|^{2}_2 \nonumber\\
& \hskip-0.08in =   & \hskip-0.08in
  \sup_{1\le m\le N} (P_0(\lambda_m))^2+(Q_0(\lambda_m))^2.
\end{eqnarray}
Set $R_n(t)= (1-t/2)^{2n} + (t/2)^{2n}, n\ge 1$. Then
 \begin{equation}\label{splinefilterbank.thm.pf.eq2}
 \inf_{0\le t\le 2} R_n(t)= 2^{-2n+1}\ {\rm and}\  \sup_{0\le t\le 2} R_n(t)=1.
 \end{equation}
Taking $P_0(t)=(1-t/2)^n$ and $P_1(t)=(t/2)^n$  in \eqref{splinefilterbank.thm.pf.eq3}
 and  applying \eqref{splinefilterbank.thm.pf.eq2}, we get
\begin{eqnarray}\label{splinefilterbank.thm.pf.eq3++}
2^{-2n+1} \|{\bf x}\|_2^2 & \hskip-0.08in \le   & \hskip-0.08in {\bf x}^T\big( ( {\bf H}_{0, n}^{\rm spln})^T {\bf H}_{0,n}^{\rm spln} + ({\bf H}_{1, n}^{\rm spln})^T  {\bf H}_{1, n}^{\rm spln}\big) {\bf x}\nonumber\\
& \hskip-0.08in =  & \hskip-0.08in   \|{\bf H}_{0, n}^{\rm spln} {\bf x}\|^{2}_2+\|{\bf H}_{1, n}^{\rm spln} {\bf x}\|^{2}_2
\le \|{\bf x}\|_2^2
\end{eqnarray}
for all ${\bf x}\in \ell^2$.
Therefore spline  filter banks
 $({\bf H}_{0, n}^{\rm spln}, {\bf H}_{1, n}^{\rm spln})$ of order $n\ge 1$ satisfy Assumption \ref{analysis.assumption3} with
 lower bound $2^{-n+1/2}$ and upper bound $1$ by
 \eqref{splinefilterbank.thm.pf.eq3} and \eqref{splinefilterbank.thm.pf.eq3++}.

Filters in a stable filter bank on $\ell^p$ are graph filters on $\ell^p, 1\le p\le \infty$.
In the following theorem, we show that analysis filter banks are stable on $\ell^p, 1\le p\le \infty$, with
quantitative estimates on their lower and upper  stability bounds by some constants  independent of the size of the graph.

\begin{theorem} \label{filterbankpnorm.thm} \rm  Let
 ${\mathcal G}$ be a  graph satisfying Assumptions \ref{assumption1} and \ref{assumption2},
  ${\bf H}_0$ and ${\bf H}_1$ have bandwidth $\sigma\ge 1$, and set ${\bf H}:={\bf H}_0^T {\bf H}_0+{\bf H}_1^T{\bf H}_1$.
If  $({\bf H}_0, {\bf H}_1)$ is stable on $\ell^2$, then
it is stable on $\ell^p$ for all $1\le p\le \infty$. Moreover,  we have the following estimates
 for its lower and upper  stability bounds $C_p$ and  $D_p$:
 \begin{equation}\label{filterbanknorm.thm.eq1}
 C_p\ge  \frac{\|{\bf H}\|_{{\mathcal B}_2}^{1/2}}
 {d!  2^{d+1}  ( D_1({\mathcal G}))^{2}  (\sigma+1)^{2d} \kappa^{d+2}}
  \end{equation}
 and
 \begin{equation}  \label{filterbanknorm.thm.eq2}
 D_p \le  2 D_1({\mathcal  G}) (\sigma+1)^d   \|{\bf H}\|_{{\mathcal B}_2}^{1/2},
 \end{equation}
 where $d$ and $D_1({\mathcal G})$ are the
Beurling dimension and density of the graph ${\mathcal G}$ respectively, and
 \begin{equation}\label{kappa.def}
\kappa=\|{\bf H}^{-1}\|_{{\mathcal B}_2} \|{\bf H}\|_{{\mathcal B}_2}
\end{equation}
is the condition number of the matrix ${\bf H}$.
\end{theorem}

 Combining \eqref{splinefilterbank.thm.pf.eq3++}
 and   Theorem \ref{filterbankpnorm.thm}, the spline
 filter banks
  $({\bf H}_{0, n}^{\rm spln}, {\bf H}_{1, n}^{\rm spln}), n\ge 1$, are stable on $\ell^p, 1\le p\le \infty$, and
their lower and upper stability bounds $C_p$ and  $D_p$ satisfy
   \begin{eqnarray*}\label{splinefilterbanknorm.cor.eq1}
\frac{1} { 2^{2(d+2)n-1} d!     (n+1)^{2d} (D_1({\mathcal  G}))^{2}} &\hskip-0.08in \le &\hskip-0.08in C_p\le D_p\nonumber\\
&\hskip-0.08in   \le   & \hskip-0.08in 2 D_1({\mathcal  G})  (n+1)^{d+1}. \end{eqnarray*}

We finish this section  with a remark on stability bounds of a graph filter bank on the space $\ell^2$ and on the  spaces $\ell^p, p\ne 2$.

\begin{remark} \rm
 For a finite graph ${\mathcal G}=(V, E)$, a stable filter bank  $({\bf H}_0, {\bf H}_1)$
on $\ell^2$ is also stable on $\ell^p, 1\le p\le \infty$, and   the  lower   stability bounds  $C_2$ and $C_p$  satisfy
\begin{equation}\label{Mstability}
N^{-|1/p-1/2|}\le \frac{C_2}{C_p}\le  N^{|1/p-1/2|},
\end{equation}
where $N=\# V$ is the size of the graph ${\mathcal G}$. The above estimation is unfavorable
when the graph ${\mathcal G}$ has large size,
 however it cannot be improved  if there is no restriction on the  filter bank $({\bf H}_0, {\bf H}_1)$.
As our analysis filter bank $({\bf H}_0, {\bf H}_1)$ has small bandwidth $\sigma$,
we obtain the following estimate independent of the size  $N$ of the graph ${\mathcal G}$ from  Proposition \ref{filterbank.prop} and Theorem  \ref{filterbankpnorm.thm},
 \begin{equation}\hskip-0.15in
 \frac{1}{2 D_1({\mathcal G})  (\sigma+1)^d \kappa^{1/2}} \le \frac{C_2}{C_p}  \le
  d!  2^{d+1}  ( D_1({\mathcal G}))^{2}  (\sigma+1)^{2d} \kappa^{d+3/2},
 \end{equation}
 where  $\kappa$ is given in \eqref{kappa.def}.
The reader may refer to  \cite{cheng_SDS16} and \cite{akramjfa09}--\cite{shinsun17}
for historical remarks and various estimates on the ratio between stability bounds on $\ell^p$ and $\ell^q, 1\le p, q\le \infty$,
for  matrices with certain off-diagonal decay.
\end{remark}


\section{Synthesis filter banks and Bezout identity}  
\label{perfectreconstruction.section1}

Let ${\mathcal G}=(V, E)$  be a graph satisfying Assumptions \ref{assumption1} and \ref{assumption2},
and $({\bf H}_0, {\bf H}_1)$ be a  graph filter bank satisfying Assumptions \ref{analysis.assumption1}, \ref{analysis.assumption2} and \ref{analysis.assumption3}. In this section, we propose  an algebraic method to construct graph filters ${\bf G}_0$ and ${\bf G}_1$ so that
the NSGFB with the analysis filter bank $({\bf H}_0, {\bf H}_1)$ and synthesis filter bank $({\bf G}_0, {\bf G}_1)$
satisfies the perfect reconstruction condition \eqref{biorthogonal.condition}
and the bandwidth of synthesis filter bank $({\bf G}_0, {\bf G}_1)$ is no larger
than the bandwidth of the analysis filter bank  $({\bf H}_0, {\bf H}_1)$.
The proposed approach  applies for
 filter banks $({\bf H}_0, {\bf H}_1)$ being polynomials of the symmetric normalized Laplacian on the graph ${\mathcal G}$, i.e.,
 \begin{equation}  \label{bezout.thm.eq1}
 {\bf H}_0= P_0({\bf L}_{\mathcal G}^{\rm sym}) \ {\rm and} \ {\bf H}_1= P_1({\bf L}_{\mathcal G}^{\rm sym})
 \end{equation}
 for some polynomials $P_0$ and $P_1$.

\begin{theorem}\label{bezout.thm}\rm
Let ${\mathcal G}$  be a graph satisfying Assumptions \ref{assumption1} and \ref{assumption2},
$({\bf H}_0, {\bf H}_1)$ be a graph filter bank satisfying Assumptions \ref{analysis.assumption1}, \ref{analysis.assumption2} and \ref{analysis.assumption3} and being of the form \eqref{bezout.thm.eq1}, and
let   $0\le \lambda_1\le \lambda_2\le\cdots\le \lambda_N\le 2$ be eigenvalues  of the symmetric normalized Laplacian   ${\bf L}_{\mathcal G}^{\rm sym}$.
 If polynomials $Q_{0}$ and $Q_{1}$ satisfy
  \begin{equation}\label{bezout.thm.eq3}
  P_0(\lambda_m) Q_{0}(\lambda_m)+P_1(\lambda_m)Q_{1} (\lambda_m)=1,\ 1\le m\le N,
 \end{equation}
then
  the  NSGFB with the analysis filter bank $({\bf H}_0, {\bf H}_1)$ and  synthesis filter bank $({\bf G}_{0}, {\bf G}_{1})$
 satisfies the perfect reconstruction condition \eqref{biorthogonal.condition},
where
  \begin{equation}\label{bezout.thm.eq2}
 {\bf G}_{0}= Q_{0}({\bf L}_{\mathcal G}^{\rm sym})\ \ {\rm and} \ \ {\bf G}_{1}= Q_{1}({\bf L}_{\mathcal G}^{\rm sym}).
 \end{equation}
\end{theorem}

 The
filter ${\bf G}_{0}$ in  \eqref{bezout.thm.eq2}  passes the
 normalized constant signal  ${\bf D}_{\mathcal G}^{1/2} {\bf 1}$, since
${\bf G}_{0} {\bf D}_{\mathcal G}^{1/2} {\bf 1}=Q_{0}(0){\bf D}_{\mathcal G}^{1/2} {\bf 1}= {\bf D}_{\mathcal G}^{1/2} {\bf 1}$,
where the last equation follows from \eqref{p0p1.eq} and \eqref{bezout.thm.eq3}.
However,  the
filter ${\bf G}_{1}$ in  \eqref{bezout.thm.eq2} may not block the
 normalized constant signal, as
 ${\bf G}_{1} {\bf D}_{\mathcal G}^{1/2} {\bf 1}=Q_{1}(0){\bf D}_{\mathcal G}^{1/2} {\bf 1}$ is not necessarily a zero signal.
In this case, we can construct a new synthesis filter bank by lifting, 
\begin{equation}\label{splinesynthesis.eq3}
 \widetilde {\bf G}_{0}= {\bf G}_{0}+Q_{1}(0) {\bf H}_1\ {\rm and} \ \widetilde {\bf G}_{1}= {\bf G}_{1}-Q_{1}(0) {\bf H}_0,
 \end{equation}
 which satisfies
 $\widetilde {\bf G}_{0} {\bf D}_{\mathcal G}^{1/2} {\bf 1}= {\bf D}_{\mathcal G}^{1/2} {\bf 1}$ and  $
 \widetilde {\bf G}_{1} {\bf D}_{\mathcal G}^{1/2} {\bf 1}={\bf 0}$.

  A strong version of \eqref{bezout.thm.eq3} is the Bezout identity
 \begin{equation}\label{Bezout.identity}
 P_0 (z)Q_{0}(z)+ P_1(z) Q_{1}(z)=1, \ z\in \CC
 \end{equation}
 for polynomials $P_0, P_1, Q_{0}$ and $Q_{1}$. In the circulant graph setting, the above approach of constructing synthesis
 filter banks via solving Bezout identity \eqref{Bezout.identity} was discussed in \cite{dragotti2017}.
 Comparing with the  Bezout identity \eqref{bezout.thm.eq3} on the eigenvalue set of the symmetric normalized Laplacian matrix ${\bf L}_{\mathcal G}^{\rm sym}$,
 the advantage of the  approach  \eqref{Bezout.identity}
  provides a tool to  design synthesis filter banks without  a priori knowledge of global topology of the residing graph and then it
   simplifies the design of synthesis filter banks for signal
reconstruction.
 It is well known that the Bezout identity  \eqref{Bezout.identity}
  is solvable  if and only if polynomials $P_0$ and $P_1$ have no common root.
 Moreover, there is a unique solution pair $(Q_{0, B}, Q_{1, B})$ to the  Bezout identity  \eqref{Bezout.identity}
such that  $Q_{0, B}(0)=1$,  $Q_{1, B}(0)=0$ and   the degree of $Q_{0, B}$ (resp. $Q_{1, B}$) is no larger than the degree of $P_1$ (resp. $P_0$).  Define
 \begin{equation}  \label{splinesynthesis.thm.eq1}
 {\bf G}_{0, B}= Q_{0, B}({\bf L}_{\mathcal G}^{\rm sym}) \ \ {\rm and}\  \ {\bf G}_{1, B}= Q_{1, B}({\bf L}_{\mathcal G}^{\rm sym}).
 \end{equation}
 Then the bandwidth of the synthesis filter bank $({\bf G}_{0, B}, {\bf G}_{1, B})$ is no larger than bandwidth of the analysis filter bank
 $({\bf H}_0, {\bf H}_1)$. Moreover, for any synthesis filter bank $({\bf G}_0, {\bf G}_1)$ there exists a polynomial $R$
 such that
 \begin{equation}\label{R.selection}
 {\bf G}_0= {\bf G}_{0, B}+{ R}({\bf L}_{\mathcal G}^{\rm sym})  {\bf H}_1 \  \ {\rm  and}\ \  {\bf G}_1= {\bf G}_{1, B}- {R} ({\bf L}_{\mathcal G}^{\rm sym}) {\bf H}_0
 \end{equation}
 satisfies the perfect reconstruction condition \eqref{biorthogonal.condition}.
We remark that the  above polynomials $R$  could be  appropriately chosen for  real world applications of an NSFGB.

Following \eqref{splinesynthesis.thm.eq1}, we define {\em synthesis spline filters}  ${\bf G}_{0, n}^{\rm B, spln}$ and ${\bf G}_{1, n}^{\rm B, spln}$ of order $n\ge 1$ by
 \begin{equation}  \label{splinesynthesis.eq00}
 {\bf G}_{0, n}^{\rm B, spln}= Q_{0, n}^{\rm B, spln}({\bf L}_{\mathcal G}^{\rm sym}) \ \ {\rm and}\  \ {\bf G}_{1, n}^{\rm B, spln}= Q_{1, n}^{\rm  B,  spln}({\bf L}_{\mathcal G}^{\rm sym}),
 \end{equation}
 where  
\begin{eqnarray*} Q_{0, n}^{\rm B, spln}(t) & \hskip-0.08in = & \hskip-0.08in  \sum_{l=0}^{n-1} {{2n-1}\choose {l}} \Big(1-\frac{t}{2}\Big)^{n-1-l} \Big(\frac{t}{2}\Big)^l\\
& \hskip-0.08in & \hskip-0.08in  + {{2n-1}\choose {n-1}} \Big(\frac{t}{2}\Big)^n
\end{eqnarray*}
and
\begin{eqnarray*} Q_{1, n}^{\rm B,  spln}(t) & \hskip-0.08in = & \hskip-0.08in  \sum_{l=0}^{n-1} {{2n-1}\choose {l}} \Big(\frac{t}{2}\Big)^{n-1-l} \Big(1-\frac{t}{2}\Big)^l\\
& \hskip-0.08in & \hskip-0.08in  - {{2n-1}\choose {n-1}} \Big(1-\frac{t}{2}\Big)^n.
\end{eqnarray*}
For $n\ge 1$, the filter ${\bf G}_{0, n}^{\rm B, spln}$   passes the
 normalized constant signal  ${\bf D}_{\mathcal G}^{1/2} {\bf 1}$,
 the filter ${\bf G}_{1, n}^{\rm B, spln}$   blocks the
 normalized constant signal  ${\bf D}_{\mathcal G}^{1/2} {\bf 1}$,
 and  the NSGFB with the analysis spline  bank  $({\bf H}_{0, n}^{\rm spln}, {\bf H}_{1, n}^{\rm spln})$ and  synthesis filter bank $({\bf G}_{0, n}^{\rm  B, spln}, {\bf G}_{1, n}^{\rm B, spln})$
 satisfies the perfect reconstruction condition \eqref{biorthogonal.condition}.
The first two results follow from $Q_{0, n}^{\rm B, spln}(0)=1$ and $Q_{1, n}^{\rm B, spln}(0)=0$, while  the perfect reconstruction conclusion holds since
\begin{eqnarray*}
& \hskip-0.08in  & \hskip-0.08in \Big(1-\frac{t}{2}\Big)^n Q_{0, n}^{\rm B, spln}(t)+ \Big(\frac{t}{2}\Big)^n  Q_{1, n}^{\rm B, spln}(t) \nonumber\\
& \hskip-0.08in = & \hskip-0.08in
 \sum_{l=0}^{n-1} {{2n-1}\choose {l}} (1-u)^{2n-1-l} u^l\nonumber\\
& \hskip-0.08in  & \hskip-0.08in   +
 \sum_{l=0}^{n-1} {{2n-1}\choose {l}}    (1-u)^{l}  u^{2n-1-l}
 \nonumber\\
& \hskip-0.08in = & \hskip-0.08in ( (1-u)+u)^{2n-1}=1,
\end{eqnarray*}
where $u=t/2$.
%

In real world applications of an NSGFB such as the proposed distributed denoising in Section \ref{simulations.section},
the subband signals  ${\bf z}_0$ and ${\bf z}_1$ in \eqref{analysisoutput} are processed via some (non)linear
procedure, such as hard/soft thresholding  
 and quantization.
In this case, the reconstructed signal $\tilde {\bf x}$ is not necessarily the same as the original signal ${\bf x}$. In the following theorem, we show that
the difference is mainly dominated by the error caused by the subband processing.

\begin{proposition}\label{bezout.error.pr} \rm  Let the graph ${\mathcal G}$, the analysis filter bank $({\bf H}_0, {\bf H}_1)$ and the synthesis filter bank
$ (\mathbf{G}_{0},\mathbf{G}_{1})$ be as in Theorem \ref{bezout.thm}. 
Assume that
the error  caused by the subband processing  $\Psi_l$ on subband signals  ${\bf z}_l={\bf H}_l {\bf x}, l=0, 1$, is  dominated by $\epsilon$ for any input signal ${\bf x}\in \ell^p$, i.e.,
 \begin{equation}\label{bezout.error.pr.eq1}
\|{\bf z}_l- \Psi_l({{\bf z}_l})\|_p \le \epsilon, \ l=0, 1,
 \end{equation}
where $\epsilon\ge 0$ and $1\le p\le \infty$.
For the input signal ${\bf x} \in \ell^p$,
  the reconstructed signal
   $\tilde {\bf x}= {\bf G}_0\Psi_0({\bf z}_0)+ {\bf G}_1 \Psi_1({\bf z}_1)$
      via the corresponding NSGFB belongs to $\ell^p$ as well. Moreover
 \begin{equation}\label{bezout.error.pr.eq2}
 \|\tilde {\bf x}-{\bf x}\|_p\le   D_1({\mathcal G}) (\tilde \sigma+1)^d (\|{\bf G}_0\|_\infty+ |{\bf G}_1\|_\infty) \epsilon,
  \end{equation}
where  $d$ and $D_1({\mathcal G})$ are the
Beurling dimension and density of the graph ${\mathcal G}$ respectively, and $\tilde \sigma$ is the bandwidth of
the synthesis  filter bank (${\bf G}_0, {\bf G}_1)$.
 \end{proposition}

We finish this section with a distributed implementation of
the NSGFB with analysis/synthesis filter banks selected
in Theorem \ref{bezout.thm}.
Write  ${\bf G}_l=(g_l(i,j))_{i,j\in V}, l=0, 1$.
As the synthesis filters ${\bf G}_0$ and ${\bf G}_1$ have finite bandwidth  $\tilde \sigma$,
the synthesis procedure can be implemented in a distributed manner,
\begin{equation}\label{bezout.synthesis.implementation}
\tilde x_k=\sum_{\rho(j, k)\le \tilde \sigma}
(g_0(k,j) \tilde z_0(j)+ g_1(k,j) \tilde z_1(j)), \ k\in V,\end{equation}
where  $\tilde {\bf x}=(\tilde x_i)_{i\in V}$ is the reconstructed signal and $\Psi_l({\bf z}_l)=(\tilde z_l(i))_{i\in V}, l=0, 1$, are outputs of subband processing.
 Hence values of the reconstructed signals $\tilde {\bf x}$  at each vertex $k\in V$
  are weighted sums of values of the subband processed outputs $\Psi_0({\bf z}_0)$ and $\Psi_1({\bf z}_1)$ in a $\tilde \sigma$-neighborhood of $k \in V$, cf. \eqref{analysis.implementation} for distributed implementation of the analysis procedure.

Our representative  subband processing procedures  $\Psi$  are hard(soft) thresholding and uniform  quantization.
For those cases, the subband processing  $\Psi$ is of the form
$\Psi({\bf z})= (\psi(z_i))_{i\in V}  $
for ${\bf z}=(z_i)_{i\in V}$, where $\psi$ is the hard(soft) thresholding and uniform  quantization function.
Thus the subband processing can be implemented in a distributed manner
and the error resulted are bounded (i.e., \eqref{bezout.error.pr.eq1} holds for $p=\infty$) by the hard(soft) thresholding and   quantization level.
This together with  \eqref{analysis.implementation} and
\eqref{bezout.synthesis.implementation} implies that
the NSGFB with analysis/synthesis filter banks
in Theorem \ref{bezout.thm} can be  implemented in a distributed manner too, provided that
the subband processing  can be.

\section{Synthesis filter bank and optimization}
\label{synthesisfilterbank.section2}

Let ${\mathcal G}=(V, E)$  be a graph satisfying Assumptions \ref{assumption1} and \ref{assumption2},
and $({\bf H}_0, {\bf H}_1)$ be a graph filter bank satisfying Assumptions \ref{analysis.assumption1}, \ref{analysis.assumption2} and \ref{analysis.assumption3}.
In this section, we consider the construction of   synthesis filter banks  $(\mathbf{G}_{0}, {\bf G}_1)$
 of an NSGFB by solving the minimization problem:
\begin{equation} \label{frobenius_optimization}
\underset{\mathbf{G}_{0},\mathbf{G}_{1}}{\text{minimize}}\
 \|\mathbf{G}_{0}\|^{2}_{F}+\| \mathbf{G}_{1}\|^{2}_{F}
\end{equation}
subject to
the perfect reconstruction  condition
\begin{equation}\label{frobenius_optimization2}
\mathbf{G}_{0}\mathbf{H}_{0}+\mathbf{G}_{1}\mathbf{H}_{1}=\mathbf{I}.
\end{equation}

Define the Lagrange function  $\mathcal L$ of the constrained optimization problem  \eqref{frobenius_optimization}  and
\eqref{frobenius_optimization2} by
\begin{eqnarray*} {\mathcal L}({\bf G_0}, {\bf G_1}, {\pmb \Theta}) & \hskip-0.08in =  & \hskip-0.08in \| {\bf G}_0\|_F^2 +\|{\bf G}_1\|_F^2\\
& \hskip-0.08in  & -  \operatorname{tr}\big(
({\bf G}_0 {\bf H}_0+{\bf G}_1 {\bf H}_1- {\bf I})\pmb  \Theta^T\big).\end{eqnarray*}
By direct calculation, we have
\begin{equation}
\left\{\begin{array}{l}
\frac{\partial {\mathcal L}}{\partial {\bf G}_0}= 2 {\bf G}_0-\pmb \Theta {\bf H}_0^T\\
\frac{\partial {\mathcal L}}{\partial {\bf G}_1}= 2 {\bf G}_1-\pmb \Theta {\bf H}_1^T\\
\frac{\partial {\mathcal L}}{\partial {\bf \Theta}}={\bf G}_0 {\bf H}_0+{\bf G}_1 {\bf H}_1- {\bf I}.
\end{array}\right.
\end{equation}
Set ${\bf H}=\mathbf{H}^{T}_{0} \mathbf{H}_{0}+\mathbf{H}^{T}_{1} \mathbf{H}_{1}$.
Solving
$$\frac{\partial  {\mathcal  L}}{\partial {\bf G}_0}= \frac{\partial {\mathcal L}}{\partial {\bf G}_1} = \frac{\partial {\mathcal L}}{\partial {\pmb \Theta}}={\bf 0}$$
leads to the unique solution of the constrained  optimization problem \eqref{frobenius_optimization} and \eqref{frobenius_optimization2},
\begin{equation} \label{pseudoinverse}
{\mathbf G}_{0, L}=\mathbf{H}^{-1}\mathbf{H}^{T}_{0}
\ \ {\rm and}\ \ \mathbf{G}_{1, L}=\mathbf{H}^{-1}\mathbf{H}^{T}_{1}.
\end{equation}
The  synthesis filter  bank  $(\mathbf{G}_{0, L},\mathbf{G}_{1, L})$  in \eqref{pseudoinverse} satisfies 
 \begin{equation*}
 {\bf G}_{0, L} {\bf H}_0+{\bf G}_{1, L}{\bf H}_1={\bf I},
 \end{equation*}
 and  the  filter ${\bf G}_{0, L}$
  passes  the  normalized constant signal $ {\bf D}_{\mathcal G}^{1/2}{\bf 1}$, since
  $${\mathbf G}_{0, L}{\bf D}_{\mathcal G}^{1/2}{\bf 1}= \mathbf{H}^{-1}(\mathbf{H}^{T}_{0}\mathbf{H}_{0}+\mathbf{H}^{T}_{1}\mathbf{H}_{1})
  {\bf D}_{\mathcal G}^{1/2}{\bf 1}={\bf D}_{\mathcal G}^{1/2}{\bf 1}.
  $$
  We remark that ${\bf G}_{1, L}$
may not block  the  normalized constant signal $ {\bf D}_{\mathcal G}^{1/2}{\bf 1}$.

For the case that ${\bf H}$ is a diagonal matrix, the synthesis filter bank
$(\mathbf{G}_{0, L},\mathbf{G}_{1, L})$ in \eqref{pseudoinverse}
has the same bandwidth as the analysis
filter bank $(\mathbf{H}_0,\mathbf{H}_1)$, 
and
\begin{equation} \label{leastsquares.diagonalassumption}
|g_{l,L}(i,j)|\le \left\{\begin{array}{ll} \|{\bf H}^{-1}\|_{{\mathcal B}_2} \|{\bf H}_l\|_\infty   &  {\rm if}\ \rho(i,j)\le \sigma\\
0 & {\rm otherwise}, \end{array}\right.
\end{equation}
where   $\mathbf{G}_{l, L}:= \big(g_{l, L}(i,j))_{i,j\in V}, l=0, 1$.

Let $\kappa$ be  the condition number of the matrix ${\bf H}$  in \eqref{kappa.def}.
It is well known that $\kappa>1$ when ${\bf H}$ is not a diagonal matrix.
For $\kappa>1$, the synthesis filter bank
$(\mathbf{G}_{0, L}, \mathbf{G}_{1, L})$ in \eqref{pseudoinverse} does not   necessarily have a small bandwidth, however it always
 has an exponential off-diagonal decay.

 \begin{theorem}\label{synthesisdecay.thm} \rm  Let   ${\mathcal G}=(V, E)$  be a graph satisfying Assumptions \ref{assumption1} and \ref{assumption2},
 $({\bf H}_0, {\bf H}_1)$ be a graph filter bank  satisfying Assumptions \ref{analysis.assumption1}, \ref{analysis.assumption2} and  \ref{analysis.assumption3},
$\kappa$ be the condition number  of the matrix  ${\bf H}:={\bf H}_0^T {\bf H}_0+{\bf H}_1^T{\bf H}_1$, and
 let $\mathbf{G}_{l, L}:= \big(g_{l, L}(i,j))_{i,j\in V}, l=0, 1$, be as in \eqref{pseudoinverse}.
 Assume that $\kappa>1$,
 then
 \begin{eqnarray}\label{g1.exponential}
\hskip-0.35in |g_{l, L}(i,j)| & \hskip-0.08in \le & \hskip-0.08in  D_1({\mathcal G})  (\sigma+1)^{d} (1-1/\kappa)^{-1/2}\nonumber\\
& \hskip-0.08in  & \hskip-0.08in \times
\|{\bf H}^{-1}\|_{{\mathcal B}_2} \|{\bf H}_l\|_\infty
\exp \Big(-\frac{\theta}{2\sigma}\rho(i,j)\Big)
\end{eqnarray}
hold for all $i,j\in V$ and $l=0, 1$,
  where   $\theta= \ln (\kappa/(\kappa-1))$,
  $\sigma\ge 1$ is the bandwidth of the analysis filter bank $({\bf H}_0, {\bf H}_1)$,
  and  $d$ and $D_1({\mathcal G})$ are the
Beurling dimension and density of the graph ${\mathcal G}$ respectively.
\end{theorem}

\begin{remark}\label{shot.remark}\rm
Agents located at some vertices may   lose data processing  ability and/or communication capability.
In that case, outputs of the analysis procedure of an NSGFB can be considered as being corrupted by  shot noise.
The exponential off-diagonal decay property in Theorem \ref{synthesisdecay.thm} implies that
the reconstructed signal suffers mainly in their neighborhood of limited size.
This means 
that the proposed NSGFB can limit the influence of shot
noise essentially to their small neighborhoods on the graph.
\end{remark}

\begin{remark}\label{resonance.remark}\rm
By the exponential off-diagonal decay property in Theorem \ref{synthesisdecay.thm},  the synthesis filters $(\mathbf{G}_{0, L}, \mathbf{G}_{1, L})$ are  filters on $\ell^p, 1\le p\le \infty$,
  \begin{eqnarray}\label{leastsquares.lpfilters}
 \|{\bf G}_{l,L}\|_{{\mathcal B}_p} & \hskip-0.08in \le & \hskip-0.08in  d! 2^d (D_1({\mathcal G}))^2 (\sigma+1)^{2d} \kappa^{d+1}(1-1/\kappa)^{-1/2}\nonumber\\
 & \hskip-0.08in  &  \times
 \|{\bf H}^{-1}\|_{{\mathcal B}_2} \|{\bf H}_l\|_\infty, \ l=0, 1.
 \end{eqnarray}
 The above conclusion with $p=\infty$ indicates that the NSGFB does not have a resonance effect.
 \end{remark}

Applying similar argument used in the proof of Proposition \ref{bezout.error.pr},
 we have

\begin{corollary}\label{leastsquares.error.cr} \rm  Let  ${\mathcal G},
 ({\bf H}_0, {\bf H}_1), (\mathbf{G}_{0, L},\mathbf{G}_{1, L})$ be as in Theorem \ref{synthesisdecay.thm},
 and let $p, \Psi_0, \Psi_1, \epsilon $ be as in Proposition \ref{bezout.error.pr}.
Assume that the input signal ${\bf x}$  of the corresponding NSGFB
 belongs to $\ell^p$,
  then the reconstructed signal   $\tilde {\bf x}= \mathbf{G}_{0, L} \Psi_0({\bf H}_0{\bf x})+ \mathbf{G}_{1, L} \Psi_1({\bf H}_1{\bf x}) $  via the NSGFB belongs to $\ell^p$
and
 \begin{eqnarray}
 \|\tilde {\bf x}-{\bf x}\|_p & \hskip-0.08in \le  & \hskip-0.08in
 d! 2^d (D_1({\mathcal G}))^2 (\sigma+1)^{2d} \kappa^{d+1}
 \|{\bf H}^{-1}\|_{{\mathcal B}_2}\nonumber\\
 & & \hskip-.5in \times (1-1/\kappa)^{-1/2}\big(\|{\bf H}_0\|_\infty +\|{\bf H}_1\|_\infty) \epsilon.
 \end{eqnarray}
 \end{corollary}

Solving the constrainted optimization program
\eqref{frobenius_optimization}
 and \eqref{frobenius_optimization2} associated with the analysis spline filter banks $({\bf H}_{0, n}^{\rm spln}, {\bf H}_{1, n}^{\rm spln})$, we obtain the synthesis spline filter bank $({\bf G}_{0, n}^{\rm L, spln}, {\bf G}_{1, n}^{\rm L, spln}), n\ge 1$,
 where
 \begin{equation}\label{splineleastsquares.def}
 {\bf G}_{l, n}^{\rm L, spln}= \Big(  \big({\bf H}_{0, n}^{\rm spln}\big)^2+ \big({\bf H}_{1, n}^{\rm spln}\big)^2\Big)^{-1} {\bf H}_{l, n}^{\rm spln},\ l=0, 1.
 \end{equation}
 The synthesis spline filters ${\bf G}_{0, n}^{\rm L, spln}$ and ${\bf G}_{1, n}^{\rm L, spln}, n\ge 1$, have full bandwidth, however they have exponential off-diagonal decay.
 Write ${\bf G}_{l, n}^{\rm L, spln}=(g_{l, n}^{\rm L, spln}(i,j))_{i,j\in V}, l=0, 1$.
 By \eqref{g1.exponential} and Theorem \ref{synthesisdecay.thm}, we obtain that
   \begin{eqnarray*}\label{g1Spline.exponential}
\hskip-0.10in |g_{l, n}^{\rm L, spln}(i,j)| & \hskip-0.08in \le & \hskip-0.08in   2^{3n-3/2} (2^{2n-1}-1)^{-1/2} (n+1)^{d} D_1({\mathcal G})  \nonumber\\
& \hskip-0.08in  & \hskip-0.08in \times
\exp \Big(-\frac{\ln (2^{2n-1}/(2^{2n-1}-1))}{2n}\rho(i,j)\Big)
\end{eqnarray*}
hold for all $i,j\in V$ and $l=0, 1$.

\begin{figure}[h] 
\begin{center}
\includegraphics[width=42mm, height=32mm]{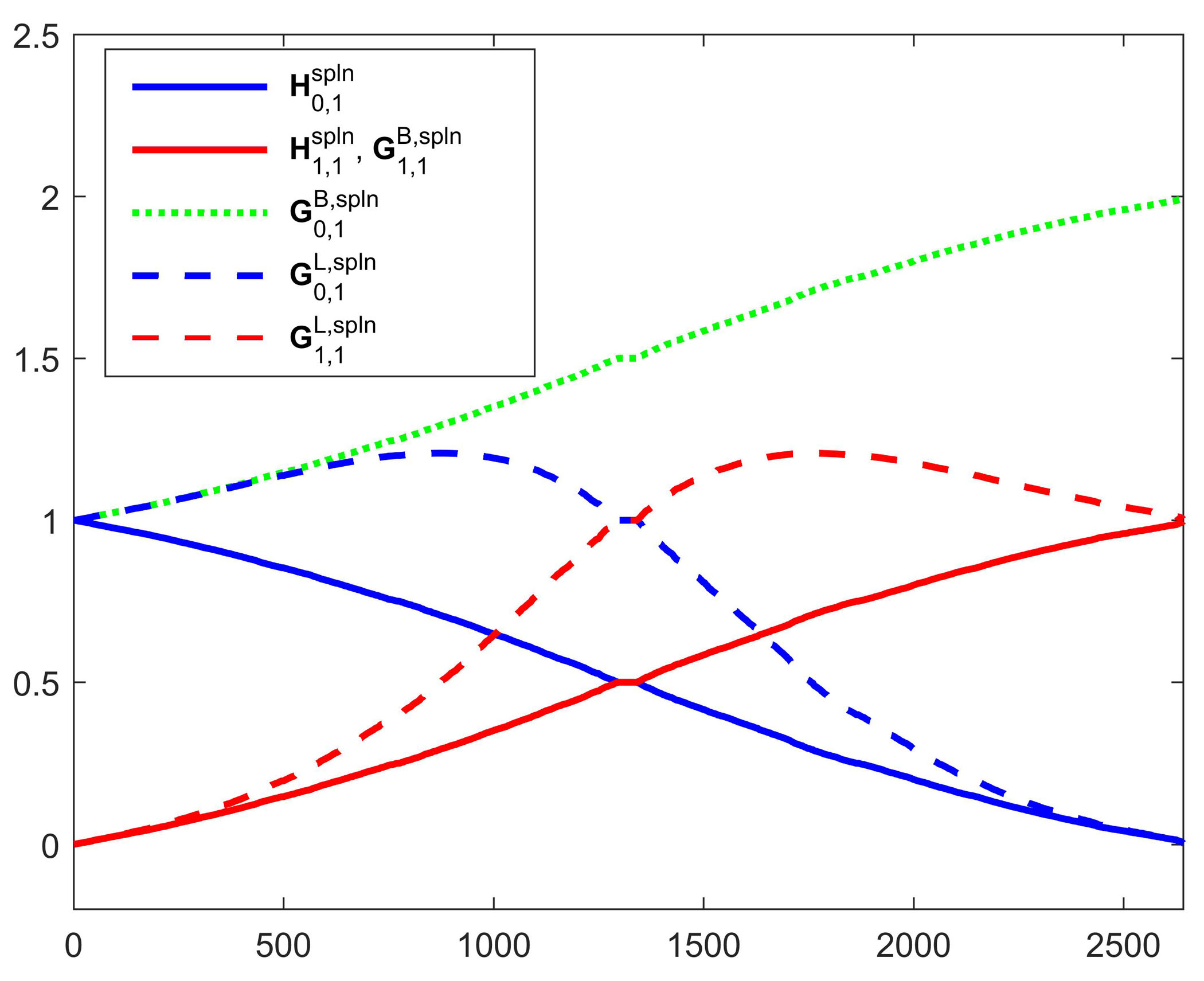}
\includegraphics[width=42mm, height=32mm]{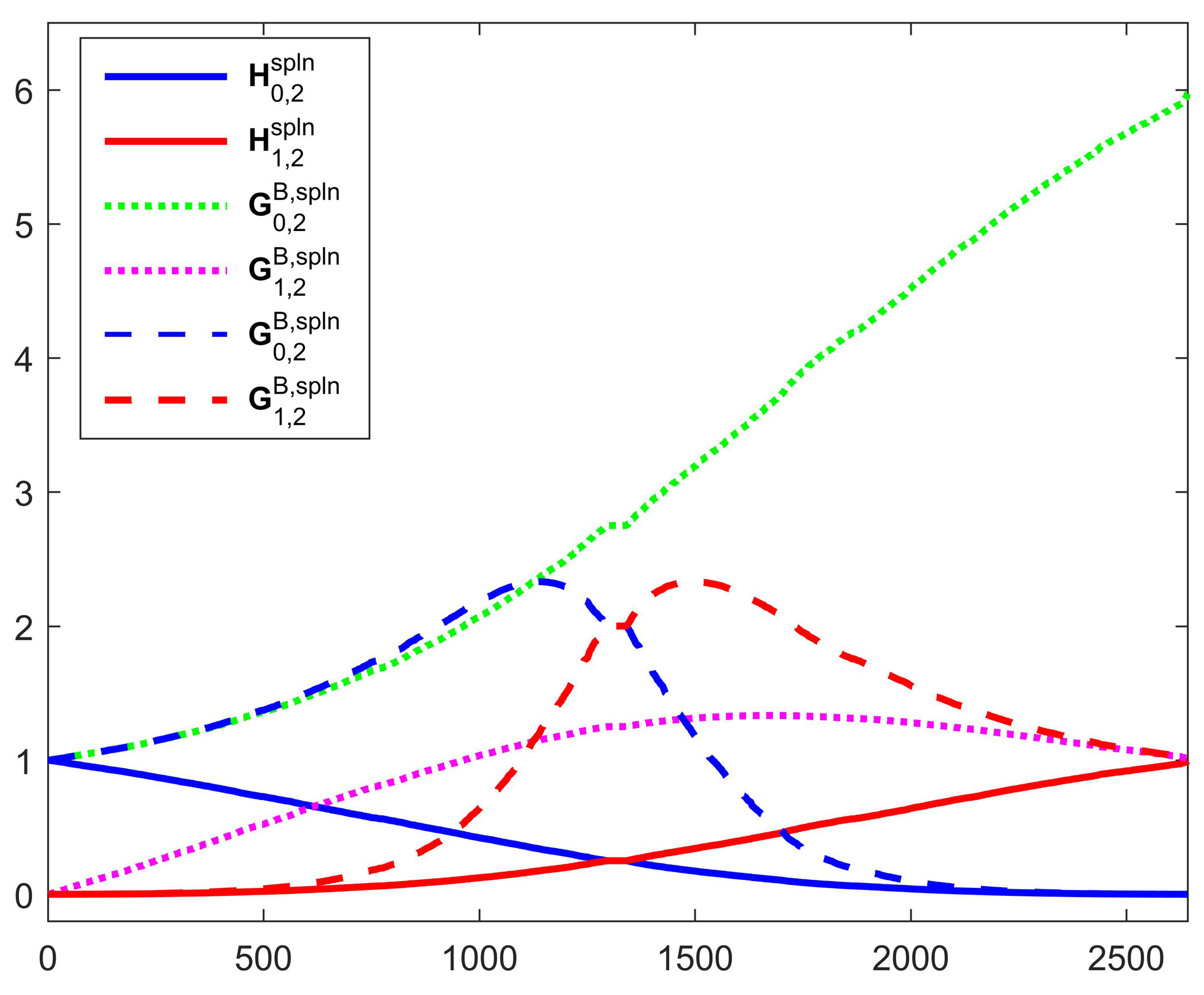}\\
\includegraphics[width=42mm, height=32mm]{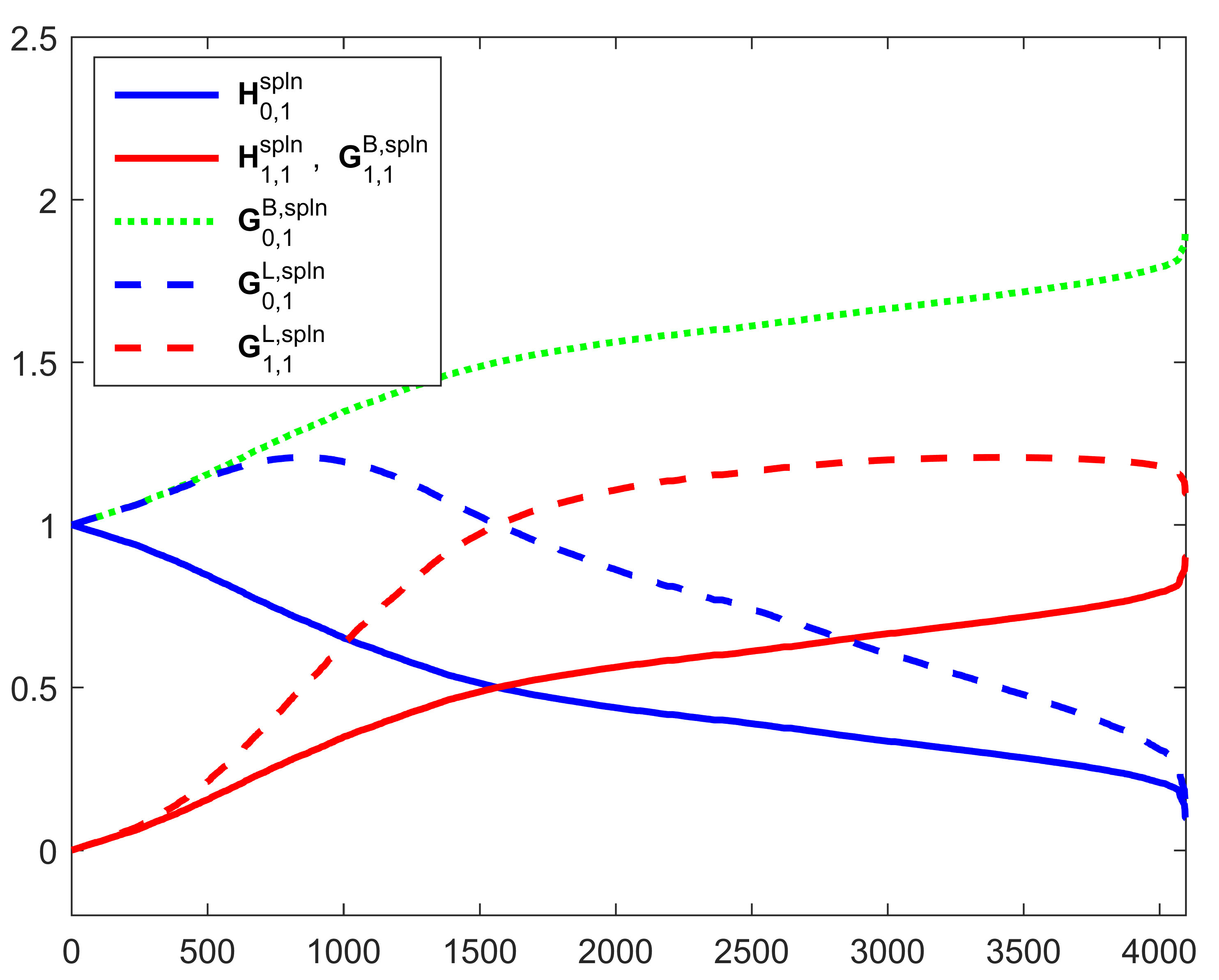}
\includegraphics[width=42mm, height=32mm]{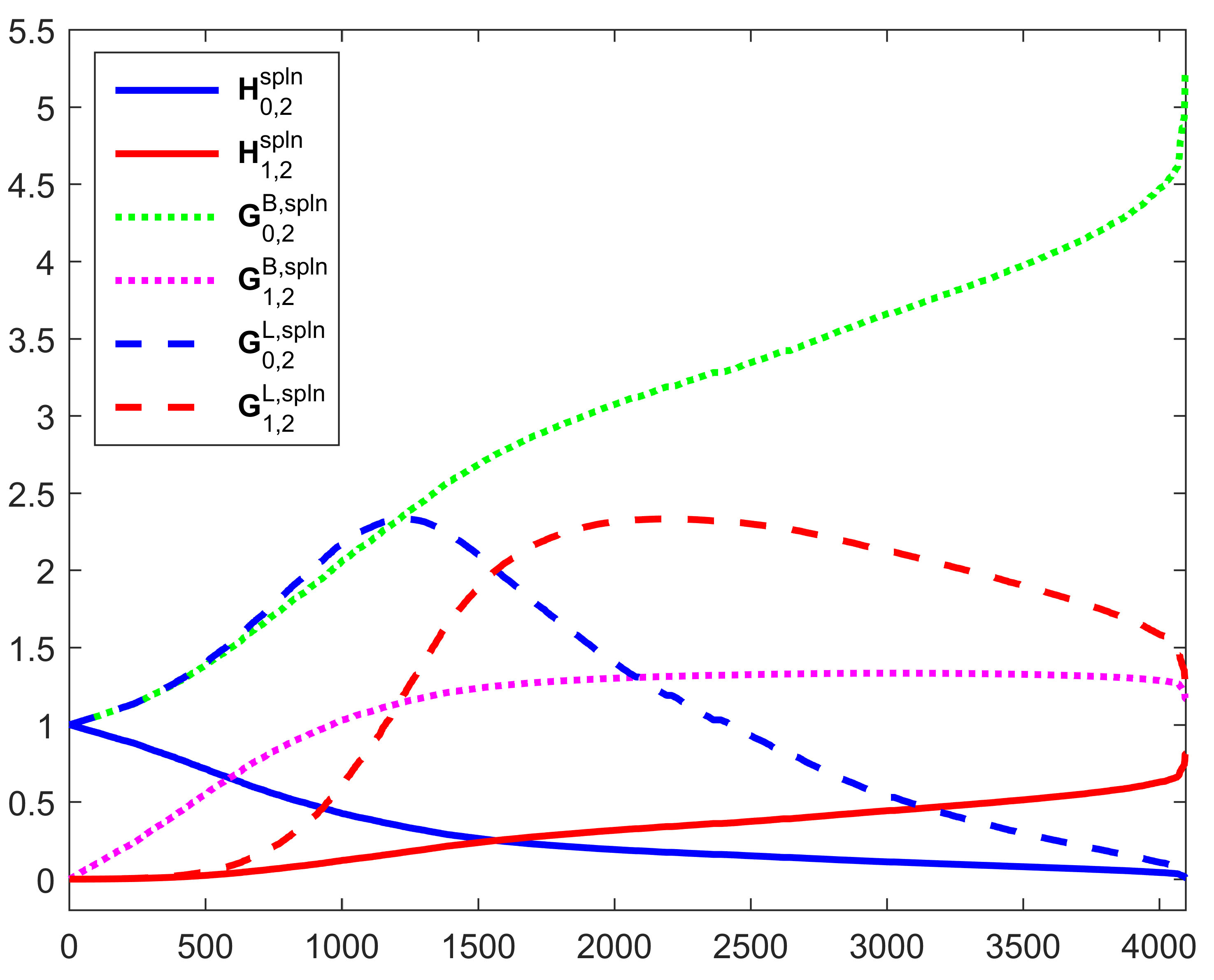}
\caption{Plotted on the top (resp. at the bottom) are the frequency responses
of   analysis/synthesis spline filters  of order $n$
 on the Minnesota traffic graph (resp. on the random geoemetric graph ${\rm RGG}_{4096}$ in Figure  \ref{circulant_graph}),
  where $n=1$ for the left figure and $n=2$ for the right figure.
     }

 \label{frequencyresponse_splinecirculant}
\end{center}
\end{figure}

 By  \eqref{polynomialfilter.eigenvalue}, we may use $P(\pmb \lambda)$
 to describe frequency response of a filter ${\bf A}=P({\bf L}_{\mathcal G}^{\rm sym})$ of the form \eqref{polynomialfilter.def},
 where  the vector
 $\pmb \lambda=(\lambda_1, \ldots, \lambda_N)$ is composed of  eigenvalues $0\le \lambda_1\le \lambda_2\le\cdots\le \lambda_N\le 2$   of the symmetric normalized Laplacian   ${\bf L}_{\mathcal G}^{\rm sym}$.
 Shown in
 Figure \ref{frequencyresponse_splinecirculant} are  frequency responses
of  the analysis spline filter banks  $({\bf H}_{0, n}^{\rm spln}, {\bf H}_{1, n}^{\rm spln})$ of order $n$,
 the  synthesis spline filter banks $({\bf G}_{0, n}^{\rm B, spln}, {\bf G}_{1, n}^{\rm  B, spln})$ in \eqref{splinesynthesis.eq00},
 and  the  synthesis spline  filter banks $({\bf G}_{0, n}^{\rm L, spln}, {\bf G}_{1, n}^{\rm  L, spln})$  just constructed, where $n=1,2$.
 It is observed that the frequency responses of  analysis spline filter banks $({\bf H}_{0, n}^{\rm spln}, {\bf H}_{1, n}^{\rm spln})$ and synthesis
 spline filter banks $({\bf G}_{0, n}^{\rm L, spln},
 {\bf G}_{1, n}^{\rm L, spln})$ have certain complementary property, while the synthesis spline filter banks $({\bf G}_{0, n}^{\rm B, spln}, {\bf G}_{1, n}^{\rm B, spln})$ constructed via solving a Bezout identity do not.

\section{Iterative distributed algorithm for synthesis procedure}
\label{ida.section}

For the NSGFB  with   synthesis filter banks 
in Theorem \ref{bezout.thm}, the distributed implementation of the corresponding synthesis procedure has been discussed in
  \eqref{bezout.synthesis.implementation}.

For the NSGFB with the analysis filter bank $({\bf H}_0, {\bf H}_1)$ and synthesis filter bank $(\mathbf{G}_{0, L},\mathbf{G}_{1, L})$
obtained from solving the constrained optimization problem \eqref{frobenius_optimization} and \eqref{frobenius_optimization2},
 output $\tilde {\bf x}$ of the synthesis procedure is
 \begin{equation}\label{leastsquares.signals}
\tilde {\bf x}= {\bf G}_{0, L} \tilde {\bf z}_0+ {\bf G}_{1, L} \tilde {\bf z}_1,
\end{equation}
where
 $\tilde {\bf z}_0$ and $\tilde {\bf z}_1$ be outputs of  subband processing.
 As filters  $\mathbf{G}_{0, L}$ and $\mathbf{G}_{1, L}$
may have full bandwidth,  it is infeasible to evaluate ${\bf G}_{0, L} \tilde {\bf z}_0$ and ${\bf G}_{1, L} \tilde {\bf z}_1$
directly in a  distributed manner.
 In this paper,   we do not intend to
find synthesis filters $\mathbf{G}_{0, L}$ and $\mathbf{G}_{1, L}$
explicitly, instead we propose an iterative distributed algorithm to implement the synthesis procedure \eqref{leastsquares.signals}.

 The proposed iterative distributed algorithm is based on two pivoting observations.
 The first observation is  that the output signal
$\tilde {\bf x}$ in \eqref{leastsquares.signals} is the unique solution of the following global least squares problem:
\begin{equation}\label{leastsquares.eq0}
\min_{{\bf x}} \ \|{\bf H}_0 {\bf x}-\tilde {\bf z}_0\|_2^2+ \|{\bf H}_1 {\bf x}-\tilde  {\bf z}_1\|_2^2,
\end{equation}
which follows from \eqref{pseudoinverse}.
To solve the  global optimization problem \eqref{leastsquares.eq0} in a distributed manner, we introduce
a family of local  least squares problems,
\begin{equation}\label{localleastsquares.eq0}
\min_{\bf x}  \|{\bf H}_0 \chi_k^{2r} {\bf x}-\tilde {\bf z}_0\|_2^2+ \|{\bf H}_1 \chi_k^{2r} {\bf x}-\tilde {\bf z}_1\|_2^2,  \ k\in V,
\end{equation}
where    $\chi^{r}_{k}, k\in V$, are  truncation operators defined by
\begin{equation}
\chi^{r}_{k}:\  (x(i))_{i\in V}\longmapsto \big( x(i)\chi_{B(k, r)}(i) \big)_{i\in V},
\end{equation}
and $r\ge 1$ is a radius parameter to be determined later \cite{cheng_SDS16}.
One may verify that given any $k\in V$, the unique solution
  of  the local optimization problem \eqref{localleastsquares.eq0}
  is given by
\begin{equation}\label{localleastsquares.solution}
{\bf v}_{k, r}
 =\chi_k^{2r} \big(\chi_k^{2r} {\bf H} \chi_k^{2r}\big)^{-1}  \chi_k^{2r} ({\bf H}_0^T \tilde {\bf z}_0+ {\bf H}_1^T \tilde {\bf z}_1),
\end{equation}
where ${\bf H}= {\bf H}_0^T{\bf H}_0+{\bf H}_1^T{\bf H}_1$.
The second  crucial observation
is that  
 the unique solution  ${\bf v}_{k, r}$
  of  the local least squares problem \eqref{localleastsquares.eq0} in the $(2r)$-neighborhood of the vertex $k$
 provides a local approximation
to the solution $\tilde {\bf x}$ of the global least squares problem \eqref{leastsquares.eq0}  in a $r$-neighborhood of the vertex $k\in V$.
Therefore we can  patch  ${\bf v}_{k, r}, k\in V$, together
\begin{equation}\label{patch_equation}
{\bf v}_r= \Big(\sum_{k'\in V}\chi_{k'}^r\Big)^{-1} \sum_{k\in V} \chi_k^r  {\bf v}_{k, r}
\end{equation}
to generate an approximation to 
the solution $\tilde {\bf x}$ of the global least squares problem \eqref{leastsquares.eq0}
in $\ell^p$ norm, i.e.,
there exists $\delta_{r, \sigma}\in (0, 1)$ such that
\begin{equation}\label{exponential.eq0}
\|{\bf v}_r- \tilde {\bf x}\|_p\le \delta_{r, \sigma} \|\tilde {\bf x}\|_p
\end{equation}
when the radius parameter $r\ge 1$ is chosen appropriately.
Set
 \begin{equation}\label{J.def}
 {\bf J}=\Big(\sum_{k' \in V}\chi_{k'}^r\Big)^{-1} \sum_{k\in V} \chi_k^r  (\chi_k^{2r} {\bf H} \chi_k^{2r})^{-1}  \chi_k^{2r}.  \end{equation}
Based on \eqref{localleastsquares.solution}, \eqref{patch_equation}   and \eqref{exponential.eq0}, we propose the following
iterative  distributed
algorithm with  initials $\tilde {\bf z}_0, \tilde {\bf z}_1\in \ell^p$:
 \begin{equation}\label{iterativedistributedalgorithm.eqm}
 \left\{\begin{array}{l}
  {\bf v}^{(m)}= {\bf J}({\bf H}_0^T \tilde {\bf z}_0^{(m-1)}+ {\bf H}_1^T \tilde {\bf z}_1^{(m-1)})\\
\tilde {\bf z}_0^{(m)}= \tilde{\bf z}_0^{(m-1)}- {\bf H}_0 {\bf v}^{(m)}\\
\tilde {\bf z}_1^{(m)}= \tilde{\bf z}_1^{(m-1)}- {\bf H}_1{\bf v}^{(m)}\\
{\bf x}^{(m)}= {\bf x}^{(m-1)}+ {\bf v}^{(m)} \end{array}\right.
\end{equation}
for $m\ge 1$, where
\begin{equation}\label{iterativedistributedalgorithm.eq0}
{\bf x}^{(0)}={\bf 0},\ \tilde {\bf z}_0^{(0)}= \tilde {\bf z}_0, \ \tilde {\bf z}_1^{(0)}=\tilde {\bf z}_1.
\end{equation}

\begin{remark}\label{jacobi.remark}\rm Decompose ${\bf H}= {\bf D}+{\bf R}$ into a diagonal component ${\bf D}$ and the remainder ${\bf R}$.
Then the classical Jacobi  method to solve the linear system
${\bf H} {\bf x}={\bf H}_0^T \tilde {\bf z}_0+ {\bf H}_1^T \tilde {\bf z}_1$
is
\begin{equation}\label{jacobi.def}
\mathbf {x} ^{(m)}={\bf D}^{-1}({\bf H}_0^T \tilde {\bf z}_0+ {\bf H}_1^T \tilde {\bf z}_1 -{\bf R}\mathbf {x} ^{(m-1)}),\  m\ge 1.\end{equation}
The above iterative method converges when ${\bf H}$ is diagonally dominated, which is not necessarily true for the case in our setting.
We observe that for $r=0$,  the matrix ${\bf J}$ in \eqref{J.def} is   equal to ${\bf D}^{-1}$. Hence the sequence ${\bf x}^{(m)}, m\ge 0$,
in the proposed algorithm \eqref{iterativedistributedalgorithm.eqm} and \eqref{iterativedistributedalgorithm.eq0} with $r=0$
is the same as the one in the Jacobi  method \eqref{jacobi.def} with initial ${\bf x}^{(0)}={\bf 0}$.
\end{remark}

Write  ${\bf H}_l=(h_l(i,j))_{i, j\in V}$  and $\tilde {\bf z}_l=(\tilde z_l(i))_{i\in V}, \ l=0, 1$. For the distributed implementation of the iterative  algorithm
\eqref{iterativedistributedalgorithm.eqm}  and \eqref{iterativedistributedalgorithm.eq0},  each agent $k\in V$ is required  to transmit  information to its neighboring vertices in $B(k, 2r+2\sigma)$, and to store the number
$m_k=\mu(B(k, r))$ of its neighboring vertices in $B(k, r)$ and four matrices ${\bf H}_{l,k}= (h_l(i,j))_{i\in B(k, 2r+\sigma), j\in B(k, 2r)}$ and
 $\widetilde {\bf H}_{l,k}= (h_l(i,j))_{i\in B(k, 2r+\sigma), j\in B(k, 2r+2\sigma)}, l=0, 1$.
Shown in Algorithm \ref{distributed_cheng.algorithm} is a
distributed implementation of the iterative  algorithm
\eqref{iterativedistributedalgorithm.eqm}  and \eqref{iterativedistributedalgorithm.eq0}, where
 every vertex
$k\in V$  is required to store  data of size $O((r+\sigma)^{2d})$,
 to
perform    $O((r+\sigma)^{2d})$  algebraic manipulations in each iteration, and
to transmit data to its  $(2r+2\sigma)$-neighborhood twice in each iteration.

\begin{algorithm}[h]
\caption{Iterative Distributed Reconstruction Algorithm}
\label{distributed_cheng.algorithm}
\begin{algorithmic}  

\STATE {\bf Inputs}:
stop criterion $\varepsilon$ and observations $\tilde {\bf z}_{l,k}=(\tilde z_l(i))_{i\in B(k, 2r+\sigma)}$ for $l=0, 1$.

\STATE {\bf Operation}:  Compute ${\bf F}_k= {\bf H}_{0,k}^T{\bf H}_{0,k}+ {\bf H}_{1,k}^T{\bf H}_{1,k}$,
 find its inverse  $({\bf F}_k)^{-1}$, and then compute  $ {\bf G}_{l,L; k}:=({\bf F}_k)^{-1} {\bf H}_{l,k}^T, l=0, 1$.

\STATE {\bf Initialization}:  ${\bf x}^{(0)}_k={\bf 0}$,  $\tilde {\bf z}_{0,k}^{(0)} =\tilde {\bf z}_{0,k}$
 and $\tilde {\bf z}_{1,k}^{(0)} =\tilde {\bf z}_{1,k}$.

\STATE{\bf Iteration}:
\STATE{\bf 1)}
 ${\bf u}_k= {\bf G}_{0,L; k}\tilde  {\bf z}_{0, k}^{(m)} + {\bf G}_{1,L; k} \tilde {\bf z}_{1, k}^{(m)} $
and write ${\bf u}_k= (u_k(i))_{i\in B(k, 2r)}$.

\STATE{\bf 2)} Communicate to all vertices $i\in B(k, r)\backslash \{k\}$ to send  data $u_k(i)$ and receive data $u_i(k)$.

\STATE{\bf 3)} Produce
$v(k)=  \frac{1}{m_k}\sum_{i\in B(k, r)} u_i(k)$.



\STATE{\bf 4)} Communicate to all vertices $i\in B(k, 2r+2\sigma)\backslash \{k\}$ to send  data $v(k)$ and receive data $v(i)$, and then generate a vector ${\bf v}_k= (v(i))_{i\in B(k, 2r+2\sigma)}$.

\STATE{\bf 5)} Update
${\bf x}^{(m+1)}_k={\bf x}^{(m)}_k+ {\bf v}_k$ and $\tilde{\bf z}_{l,k}^{(m+1)}= \tilde{\bf z}_{l,k}^{(m)}-\widetilde {\bf H}_{l, k} {\bf v}_k, l=0, 1$.

\STATE{\bf  6)} Evaluate $\|{\bf v}_{k}\|_\infty\leq \varepsilon$. If yes, terminate the iteration and output
${\bf x}^{(m+1)}_k$.
Otherwise, set $m=m+1$. 

\STATE {\bf Outputs}:  ${\bf x}^{(m+1)}_k$.  

\end{algorithmic}
\end{algorithm}

 In the next theorem, we further show that the iterative  algorithm
\eqref{iterativedistributedalgorithm.eqm} and \eqref{iterativedistributedalgorithm.eq0} converges  exponentially
 when $r$ is appropriately selected.

\begin{theorem}\label{convergence.thm} \rm  Let  $1\le p\le \infty$,   ${\mathcal G}$  be a graph satisfying Assumptions \ref{assumption1} and \ref{assumption2},
 $({\bf H}_0, {\bf H}_1)$ be a graph filter bank  satisfying Assumptions \ref{analysis.assumption1}, \ref{analysis.assumption2} and  \ref{analysis.assumption3},
$\kappa>1$ be  as \eqref{kappa.def}, the condition number  of the matrix  ${\bf H}:={\bf H}_0^T {\bf H}_0+{\bf H}_1^T{\bf H}_1$,
and
 let $(\mathbf{G}_{0, L},\mathbf{G}_{1, L})$ be as in \eqref{pseudoinverse}.
Set
\begin{equation} \label{deltarsigma.def} \delta_{r, \sigma}:= \frac{ (D_1({\mathcal G}))^2  (2\sigma+1)^d   \kappa^2}{\kappa-1}
  \exp \Big( -\frac{\theta }{2\sigma}  r\Big) (3r+2\sigma+1)^d,
\end{equation}
  where   $\theta= \ln (\kappa/(\kappa-1))$,
  $\sigma\ge 1$ is the bandwidth of the analysis filter bank $({\bf H}_0, {\bf H}_1)$,
  and  $d$ and $D_1({\mathcal G})$ are the
Beurling dimension and density of the graph ${\mathcal G}$ respectively.
  Take $\tilde {\bf z}_0, \tilde {\bf z}_1\in \ell^p$, and  let ${\bf x}^{(m)}, m\ge 0,$ be as in
\eqref{iterativedistributedalgorithm.eqm} and \eqref {iterativedistributedalgorithm.eq0}.
If the radius parameter $r$ is so chosen that
\begin{equation}\label{convergence.thm.eq2}
\delta_{r, \sigma}\in (0, 1),
\end{equation}
 then ${\bf x}^{(m)}, m\ge 0$, converges to the least squares solution $\tilde {\bf x}$ in \eqref{leastsquares.signals} exponentially,
 \begin{equation}\label{convergence.thm.eq3}
 \|{\bf x}^{(m)}-  \tilde {\bf x} \|_p\le (\delta_{r, \sigma})^m \|\tilde {\bf x}\|_p, \ m\ge 0.
 \end{equation}
\end{theorem}

\begin{remark}\rm
For $l=0, 1$, we can apply \eqref{iterativedistributedalgorithm.eqm} to prove by induction on $m$ that
$$\tilde {\bf z}_{l}^{(m)}- (\tilde {\bf z}_l-{\bf H}_l \tilde {\bf x})= -{\bf H}_0 ({\bf x}^{(m)}-\tilde {\bf x}), \ m\ge 0.$$
This together with Theorem \ref{convergence.thm} implies that
$\tilde {\bf z}_{l}^{(m)}, m\ge 1$,  in the iterative algorithm
 \eqref{iterativedistributedalgorithm.eqm}  and \eqref{iterativedistributedalgorithm.eq0} converges
 to $\tilde {\bf z}_l-{\bf H}_l \tilde {\bf x}$ exponentially, where $l=0, 1$.
\end{remark}

By \eqref{deltarsigma.def} and \eqref{convergence.thm.eq3} in Theorem \ref{convergence.thm},
 the  iterative distributed algorithm \eqref{iterativedistributedalgorithm.eqm} and
\eqref{iterativedistributedalgorithm.eq0} has fast convergence rate when a large radius parameter $r$ is chosen.
In that case,  heavier burden arises at each
iteration, which implies that each vertex in
the graph ${\mathcal G}$ should have more data storages, better computing
abilities and stronger communication capacities in real world applications.
Shown in Tables  
 \ref{minnesota_noiseless_reconstruct.newtable} and
\ref{randomsensor_noiseless_reconstruct.newtable}
are the average $E_{m,r}$ of 
$\| {\bf x}^{(m)}-{\bf x}\|_\infty/{\|{\bf x}\|_\infty}$
over 50 trials versus the number $m\ge 1$ of iterations and
the radius parameter $r\ge 0$,
where  $({\bf H}_{0, n}^{\rm spln}, {\bf H}_{1, n}^{\rm spln}), n=1, 2$ are used as  analysis filter banks,
 the signal ${\bf x}$ in Tables \ref{minnesota_noiseless_reconstruct.newtable}  and \ref{randomsensor_noiseless_reconstruct.newtable}
 is randomly selected
on the  Minnesota traffic graph and on the  random geometric graph ${\rm RGG}_{4096}$ in Figure \ref{circulant_graph} respectively.
  This demonstrates that  the iterative distributed algorithm \eqref{iterativedistributedalgorithm.eqm} and
\eqref{iterativedistributedalgorithm.eq0} converges faster for larger radius  $r$,
and the original signal can be well approximated in one step  
when
a large radius  $r$ is chosen,  see Tables
 \ref{minnesota_noiseless_reconstruct.newtable}  and \ref{randomsensor_noiseless_reconstruct.newtable}.

 By \eqref{deltarsigma.def} and Theorem \ref{convergence.thm}, there is a  radius parameter $r_0$ such that
 the  iterative distributed algorithm \eqref{iterativedistributedalgorithm.eqm} and
\eqref{iterativedistributedalgorithm.eq0} converges exponentially whenever  $r\ge r_0$.
We can select the above radius parameter $r_0$ 
 to
be independent of the size   of the graph  ${\mathcal G}$.  Our simulation indicates
that  the  iterative distributed algorithm \eqref{iterativedistributedalgorithm.eqm} and
\eqref{iterativedistributedalgorithm.eq0}  with $r=0$, i.e. the Jacobi iterative method by Remark \ref{jacobi.remark},   diverges for some bounded inputs
 on the Minnesota traffic graph and on some random geometric graphs, see the first column  of Tables  \ref{minnesota_noiseless_reconstruct.newtable}
 and  \ref{randomsensor_noiseless_reconstruct.newtable}.

\begin{table}[h] 
\centering
\caption{Performance of the  iterative distributed reconstruction algorithm
 to recover  
 signals on the Minnesota traffic graph  
 }
\label{minnesota_noiseless_reconstruct.newtable}
\begin{tabular}{|c|c|c|c|c|c|c|}
\hline
\multicolumn{6}{c}{$n=1$}\\
\hline
\diagbox{$m$}{$E_{m,r}$}{$r$}
& $0$ & $1$ & $2$ & $3$ & $4$ 
& $6$
\\
\hline
1 & $.4155$ & $.2220$ & $.0375$ & $.0160$ & $.0033$ 
 & $.0003$
\\
2 & $.1355$ & $.0238$ & $.0007$ & $.0001$ & $.0000$ 
& $.0000$
\\
3 & $.0547$ & $.0039$ & $.0000$ & $.0000$ & $.0000$  
 & $.0000$
\\
4 & $.0226$ & $.0006$ & $.0000$ & $.0000$ & $.0000$  
& $.0000$
\\
5 & $.0098$ & $.0000$ & $.0000$ & $.0000$ & $.0000$  
& $.0000$
\\
10 & $.0002$ & $.0000$ & $.0000$ & $.0000$ & $.0000$ 
& $.0000$
 \\
\hline
\hline
\multicolumn{6}{c}{$n=2$}\\
\hline
\diagbox{$m$}{$E_{m,r}$}{$r$}
& $0$ & $1$ & $2$ & $3$ & $4$ 
& $6$ \\
\hline
1 & $1.1988$ & $.6563$ & $.3187$ & $.1523$ & $.0725$ 
& $.0178$
\\
2 & $1.1662$ & $.2315$ & $.0518$ & $.0136$ & $.0029$ 
& $.0002$
\\
3 & $1.4550$ & $.1162$ & $.0125$ & $.0017$ & $.0002$  
& $.0000$
\\
4 & $1.4697$ & $.0567$ & $.0026$ & $.0002$ & $.0000$ 
 & $.0000$
\\
5 & $2.5386$ & $.0296$ & $.0006$ & $.0000$ & $.0000$  
& $.0000$
\\
7 & $4.7083$ & $.0082$ & $.0000$ & $.0000$ & $.0000$ 
& $.0000$
 \\
10 & $12.7921$ & $.0013$ & $.0000$ & $.0000$ & $.0000$ 
& $.0000$
\\
14 & $52.4168$ & $.0001$ & $.0000$ & $.0000$ & $.0000$ 
& $.0000$
\\
\hline
\end{tabular}
\end{table}

\begin{table}[h]
\centering
\caption{Performance of the  iterative distributed reconstruction algorithm
 to recover 
  signals on the random geometric graph ${\rm RGG}_{4096}$ in Figure \ref{circulant_graph} 
 }
\label{randomsensor_noiseless_reconstruct.newtable}
\begin{tabular}{|c|c|c|c|c|c|c|}
\hline
\multicolumn{6}{c}{$n=1$}\\
\hline
\diagbox{m}{$E_{m,r}$}{r}
& $0$ & $1$ & $2$ & $3$ & $4$ & $5$ 
\\
\hline
1 & $.4182$ & $.1301$ & $.0156$ & $.0035$ & $.0006$ & $.0001$ 
\\
2 & $.1963$ & $.0083$ & $.0002$ & $.0000$ & $.0000$ & $.0000$ 
\\
3 & $.1143$ & $.0008$ & $.0000$ & $.0000$ & $.0000$  & $.0000$ 
\\
4 & $.0699$ & $.0000$ & $.0000$ & $.0000$ & $.0000$  & $.0000$ 
\\
10 & $.0050$ & $.0000$ & $.0000$ & $.0000$ & $.0000$ & $.0000$ 
 \\
19 & $.0001$ & $.0000$ & $.0000$ & $.0000$ & $.0000$ & $.0000$ 
\\

\hline
\hline
\multicolumn{6}{c}{$n=2$}\\
\hline
\diagbox{m}{$E_{m,r}$}{r}
& $0$ & $1$ & $2$ & $3$ & $4$ & $5$ 
\\
\hline
1 & $1.5267$ & $.4674$ & $.1487$ & $.0437$ & $.0159$ & $.0049$ 
\\
2 & $2.8586$ & $.1098$ & $.0120$ & $.0011$ & $.0001$ & $.0000$ 
\\
3 & $6.6794$ & $.0374$ & $.0014$ & $.0000$ & $.0000$  & $.0000$ 
\\
4 & $16.4089$ & $.0121$ & $.0002$ & $.0000$ & $.0000$  & $.0000$ 
\\
5 & $40.9430$ & $.0041$ & $.0000$ & $.0000$ & $.0000$  & $.0000$ 
\\
8 & $672.8632$ & $.0002$ & $.0000$ & $.0000$ & $.0000$ & $.0000$ 
 \\
\hline
\end{tabular}
\end{table}

\section{Distributed denoising}
\label{simulations.section}

\begin{figure}[h]
\begin{center}
\includegraphics[width=50mm, height=15mm]{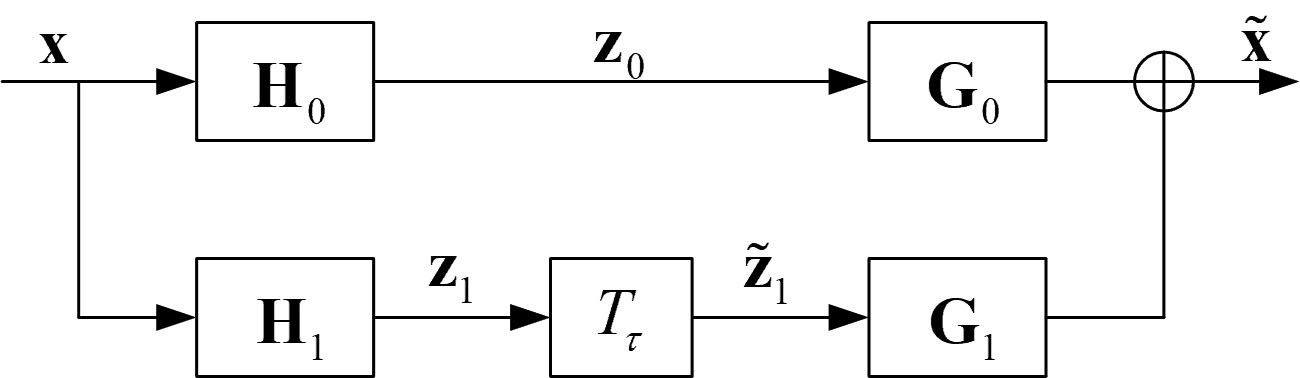} 
\caption{Block diagram of the proposed denoising procedure, 
 where  ${\bf x}$ is the noisy input  and  $\tilde {\bf x}$ is the denoised output.  }
\label{denoising_structure}
\end{center}
\end{figure}

Given an NSGFB with analysis filter bank $({\bf H}_0, {\bf H}_1)$ and synthesis filter bank $({\bf G}_0, {\bf G}_1)$, we propose
a denoising technique with hard thresholding operator $T_\tau$  applied to the high-pass subband signal, where
  $ T_\tau(t)= {\rm sgn}(t) (|t|-\tau)_+$ is the hard thresholding function
 with threshold value $\tau\ge 0$, cf. \cite{shuman13b, tremblay16, sakiyama14, sakiyama16}.  Presented in  Figure \ref{denoising_structure} is the block diagram of the proposed denoising procedure.
In this section, we demonstrate the performance of the proposal denoising procedure associated with spline NSGFBs, which can be implemented in a distributed manner.

In the simulations,  the noisy input  is
\begin{equation}\label{noisyinput.def}
{\bf x}= {\bf x}_o+{\pmb \epsilon},\end{equation}
 where  ${\bf x}_o=(x_{o, i})_{i\in V}$ is  the original graph  signal
 and the  noise ${\pmb \epsilon}=(\epsilon_i)_{i\in V}$  
has    value $\epsilon_i$ at vertex $i\in V$  randomly selected in the range $[-\eta, \eta]$.
The  spline NSGFBs   have analysis spline filter banks
 $({\bf H}_{0, n}^{\rm spln}, {\bf H}_{1, n}^{\rm spln})$ in \eqref{lowpassspline.def} and synthesis spline filter banks  being either $({\bf G}_{0, n}^{\rm B, spln}, {\bf G}_{1, n}^{\rm B, spln})$ in \eqref{splinesynthesis.eq00}
or $({\bf G}_{0, n}^{\rm L, spln}, {\bf G}_{1, n}^{\rm L, spln})$ in \eqref{splineleastsquares.def}, where $n\ge 1$.
They are abbreviated by  NSGFB-B$n$ and NSGFB-L$n$ respectively.
The  denoising procedure  is  performed by
retaining the low-pass subband  signal ${\bf z}_{0}={\bf H}_{0, n}^{\rm spln} {\bf x}$
and applying the hard thresholding operation $T_\tau$ to the high-pass subband signal $ {\bf z}_{1}={\bf H}_{1, n}^{\rm spln} {\bf x}$, where
$\tau>0$ is chosen appropriately.
Thus  the denoised output
 is
 $$\tilde {\bf x}= {\bf G}_{0, n}^{\rm B, spln} {\bf z}_{0}+ {\bf G}_{1, n}^{\rm B, spln} T_\tau({\bf z}_{1})$$
 for    NSGFB-B$n$,
 and
 $$\tilde {\bf x}= {\bf G}_{0, n}^{\rm L, spln} {\bf z}_{0}+ {\bf G}_{1, n}^{\rm L, spln} T_\tau ({\bf z}_{1})$$
 for     NSGFB-L$n$ respectively, where $n\ge 1$.
 For the above denoising procedure,  we use $20\log_{10} \|{\bf x}_o\|_p/\|{\bf x}-{\bf x}_o\|_p$
 to measure the input $\ell^p$-signal-to-noise ratio ($\ell^p$-SNR)  in dB, and
 $20\log_{10} \|{\bf x}_o\|_p/\|\tilde {\bf x}-{\bf x}_o\|_p$ to measure the output $\ell^p$-SNR in dB,
  where
 $1\le p\le \infty$.

The Minnesota traffic graph is a test bed for various techniques in signal processing on graphs of medium size (\cite{shuman13, narang12, tanaka14, shuman13b}).
The denoising performance of the proposed spline NSGFBs  on the Minnesota graph is presented in Table \ref{Denoise_minnesota2.table},
where the original signal  ${\bf x}_o$ is the blockwise
constant function  in  Figure \ref{subband_circulant.fig},
the threshold value $\tau$
 is  selected to be $3\eta$, and  the input and output $\ell^2$-SNRs are the average values over 50 trials.
Shown also in Table \ref{Denoise_minnesota2.table} are the performance comparison with
 the biorthogonal graph filter
bank (graphBior) in \cite{narang13}, the $M$-channel oversampled graph
filter bank (OSGFB) in \cite{tanaka14},
 and  the pyramid transform (PRT) in \cite{shuman13b}, where the corresponding output $\ell^2$-SNRs  are calculated from the accompanying codes in these references.
 It indicates that the  spline NSGFBs and the OSGFB 
  outperform other two methods in the small noise  scenario, 
  the spline NSGFBs have the best performance in the moderate noise environment, 
  and the PRT stands out from the rest in  the strong noisy case. 
\begin{table}
\centering
\caption{Denoising performance on the  Minnesota traffic graph measured with the standard $\ell^2$-SNR} 
 \label{Denoise_minnesota2.table}
\begin{tabular} {c|c|c|c|c|c|c}
\hline
${\eta}$ & $1/32$ &$1/16$ & $1/8$ &$1/4$ & $1/2$ &$1$ \\
\hline
{\rm Input $\ell^2$-SNR} & $34.89$  & $28.85$ & $22.83$ & $16.82$ & $10.81$ & $4.75$ \\
\hline
{\rm graphBior} & $34.43$  & $28.91$ & $24.06$ & $18.21$ & $12.79$ & $7.39$ \\

{\rm OSGFB } & $38.25$  & $32.59$ & $24.44$ & $16.70$ & $12.54$ & $4.69$ \\

{\rm PRT} & $35.31$  & $29.41$ & $23.74$ & $18.46$ & $15.45$ & $12.77$ \\

{\rm NSGFB-B1} & $37.50$  & $31.45$ & $25.43$ & $18.91$ & $13.18$ & $7.39$ \\

{\rm NSGFB-B2} & $37.25$  &  $30.74$ &  $24.95$ &  $18.53$ & $13.32$ &  $7.68$ \\

{\rm NSGFB-L1} & $38.49$  & $32.44$ & $26.42$ & $19.25$ & $13.82$ & $8.34$ \\

{\rm NSGFB-L2} & $37.25$  & $30.67$ & $24.91$ & $18.16$ & $13.33$ & $7.88$ \\
\hline
\end{tabular}
\end{table}


 Presented in
  Tables \ref{Denoise_RRG2.table} and \ref{Denoise_gnninfinity.table} are the denoising performance of spline NSGFBs
  and  the performance comparision with the graphBior in \cite{narang13}, the OSGFB in \cite{tanaka14},
   and  the PRT in \cite{shuman13b}  
   on the  random geometric graph ${\rm RGG}_N$,
  where  $N=4096$,   the original signal  ${\bf x}_o$ is the blockwise polynomial
in  Figure \ref{subband_circulant.fig},
 the threshold value $\tau$
 is  selected to be $3\eta$, and  the input and output $\ell^2$-SNRs in Table
 \ref{Denoise_RRG2.table} and the input and output $\ell^\infty$-SNRs  in Table \ref{Denoise_gnninfinity.table}
 are the average values over 50 trials.
 It is observed that  the spline NSGFBs proposed in this paper outperform the graphBior, OSBFB and PRT in small and moderate noise scenario,
 and that the spline NSGFBs have comparable  performance with the rest in the strong noisy case.
  Also from  Tables \ref{Denoise_RRG2.table} and \ref{Denoise_gnninfinity.table}, we see that
 the differences between the  input and output $\ell^p$-SNRs for $p=2, \infty$ are in some range.
This confirms
  the conclusions  in Proposition \ref{bezout.error.pr} and
Corollary \ref {leastsquares.error.cr}  
that the output noise is dominated by a multiple of the input noise.

\begin{table}
\centering
\caption{Denoising performance on the random geometric graph ${\rm RGG}_{4096}$ measured with the  standard $\ell^2$-SNR}
\label{Denoise_RRG2.table}
\begin{tabular} {c|c|c|c|c|c|c}
\hline
${\eta}$ &  $1/32$ &$1/16$ & $1/8$ &$1/4$ & $1/2$ &$1$ \\
\hline
{\rm Input  $\ell^2$-SNR} & $35.06$  & $29.04$ & $23.02$ & $17.01$ & $10.97$ & $4.95$ \\
\hline
{\rm graphBior} & $33.82$  & $28.61$ & $23.27$ & $18.20$ & $13.21$ & $8.34$ \\

{\rm OSGFB } & $31.69$  & $26.37$ & $20.79$ & $16.40$ & $13.40$ & $11.13$ \\

{\rm PRT} & $32.89$ & $27.51$ & $22.44$ & $17.70$ & $14.21$ & $11.81$ \\

{\rm NSGFB-B1} & $37.43$  & $31.40$ & $25.34$ & $19.31$ & $13.47$ & $7.62$ \\

{\rm NSGFB-B2} & $36.65$  &  $30.63$ &  $24.89$ &  $19.37$ & $13.80$ &  $8.25$ \\

{\rm NSGFB-L1} & $38.86$  & $32.87$ & $26.61$ & $20.45$ & $14.91$ & $9.40$ \\

{\rm NSGFB-L2} & $36.08$  & $29.97$ & $24.27$ & $19.16$ & $13.92$ & $9.05$ \\

\hline
\end{tabular}
\end{table}

\begin{table}
\centering
\caption{Denoising performance on the random geometric graph ${\rm RGG}_{4096}$ measured with the  $\ell^\infty$-SNR } \label{Denoise_gnninfinity.table}
\begin{tabular} {c|c|c|c|c|c|c}
\hline
${\eta}$ & $1/32$ &$1/16$ & $1/8$ &$1/4$ & $1/2$ &$1$ \\
\hline
{\rm Input $\ell^\infty$-SNR} & $34.90$  & $28.88$ & $22.85$ & $16.83$ & $10.81$ & $4.79$ \\
\hline
{\rm graphBior} & $23.45$  & $17.28$ & $11.12$ & $5.34$ & $0.32$ & $-4.00$ \\

{\rm OSGFB } & $20.59$  & $14.32$ & $6.83$ & $0.99$ & $-1.99$ & $-2.71$ \\

{\rm PRT} & $23.24$  & $17.25$ & $11.27$ & $5.43$ & $0.39$ & $-2.15$ \\

{\rm NSGFB-B1} & $31.84$  & $25.16$ & $18.71$ & $11.05$ & $6.55$ & $2.60$ \\

{\rm NSGFB-B2} & $26.86$  &  $20.34$ &  $14.67$ &  $8.32$ & $3.55$ &  $1.67$ \\

{\rm NSGFB-L1} & $29.28$  & $22.70$ & $16.19$ & $9.28$ & $4.08$ & $0.35$ \\

{\rm NSGFB-L2} & $24.66$  & $17.91$ & $12.27$ & $6.72$ & $0.48$ & $-0.52$ \\

\hline
\end{tabular}
\end{table}

\begin{figure*}[htbp] 
\centering
\includegraphics[width=40mm, height=32mm]{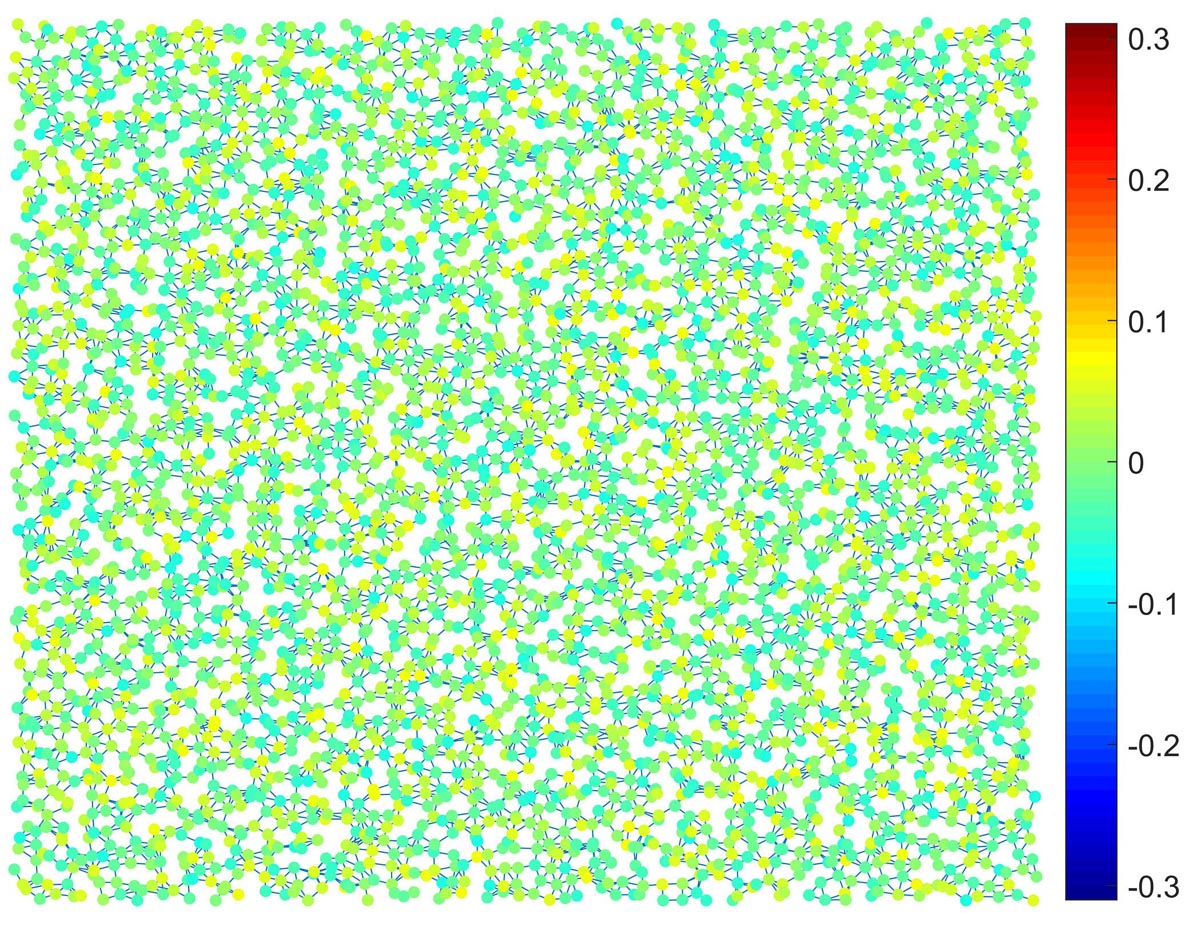}
\includegraphics[width=40mm, height=32mm]{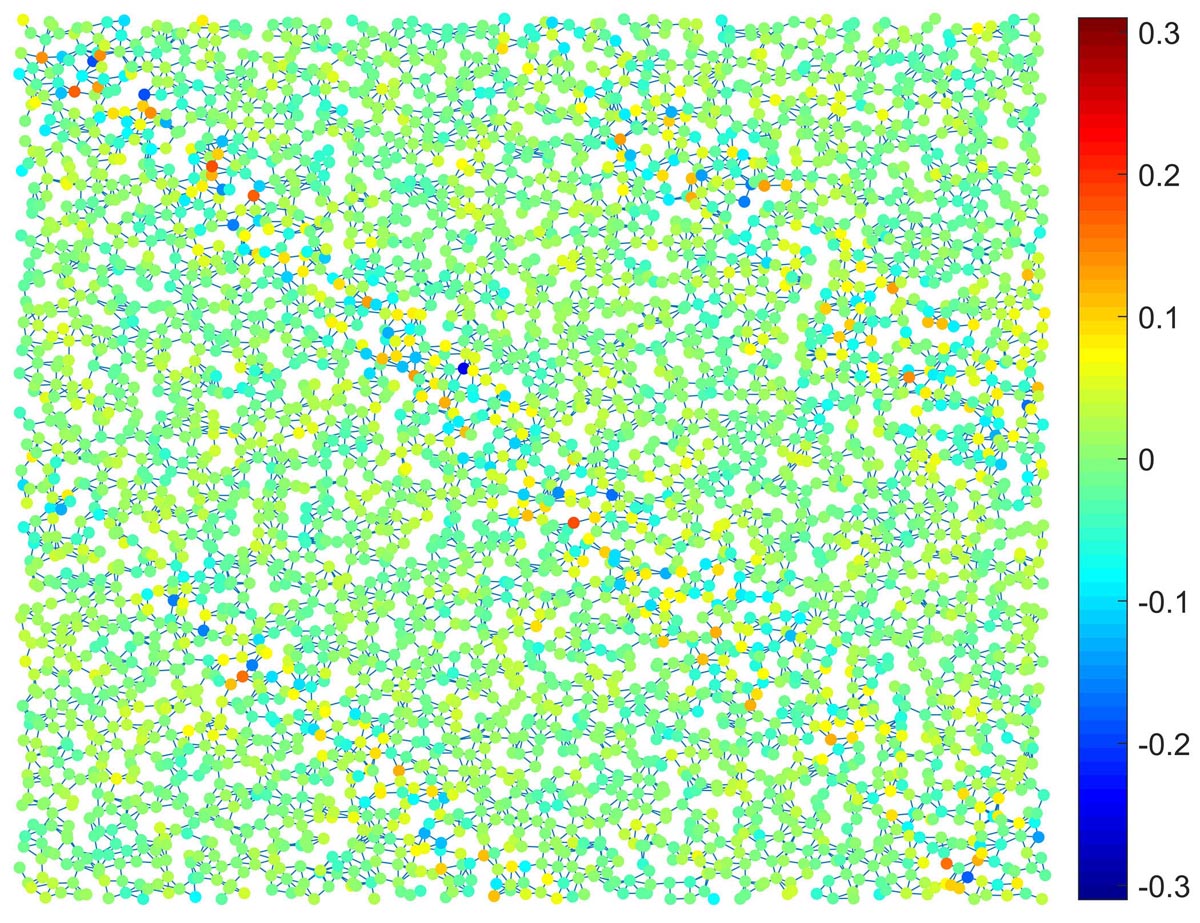}
\includegraphics[width=40mm, height=32mm]{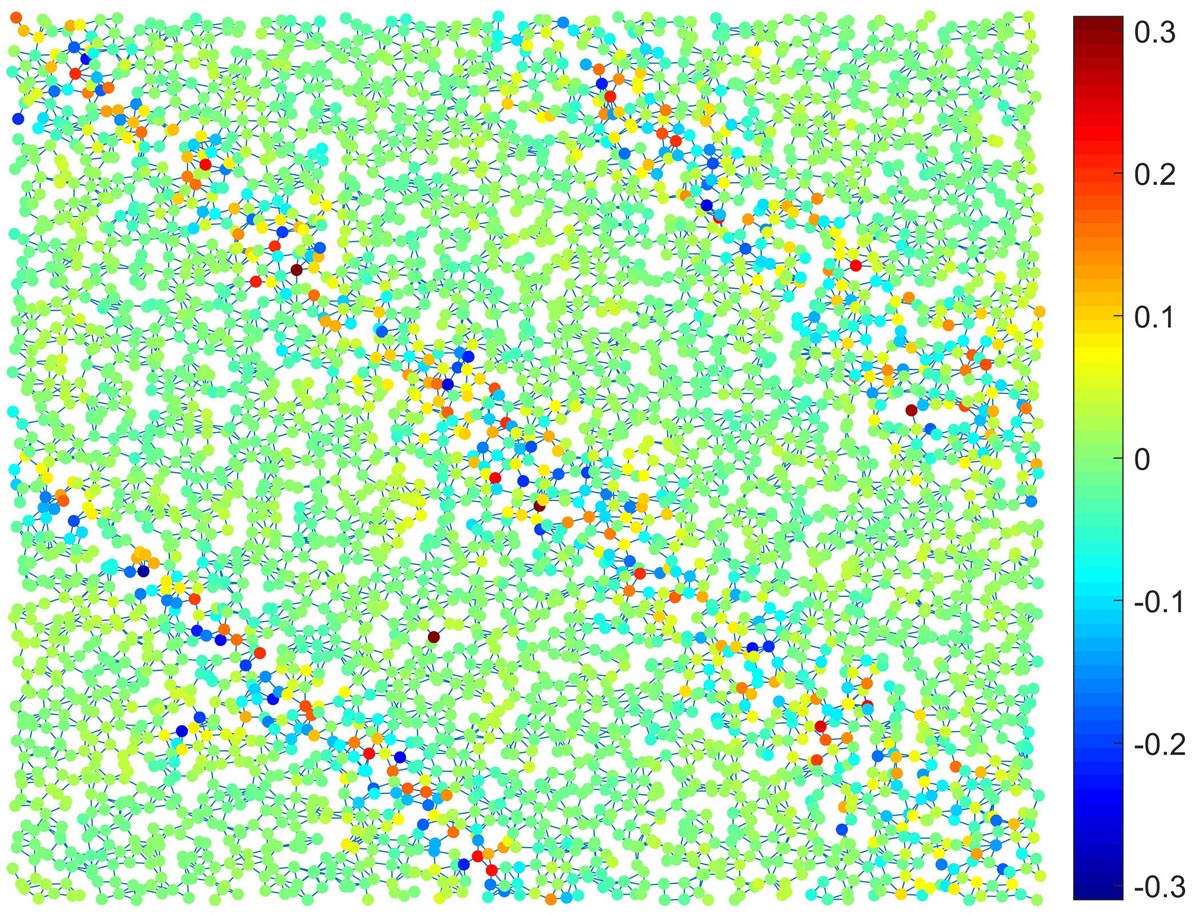}
\includegraphics[width=40mm, height=32mm]{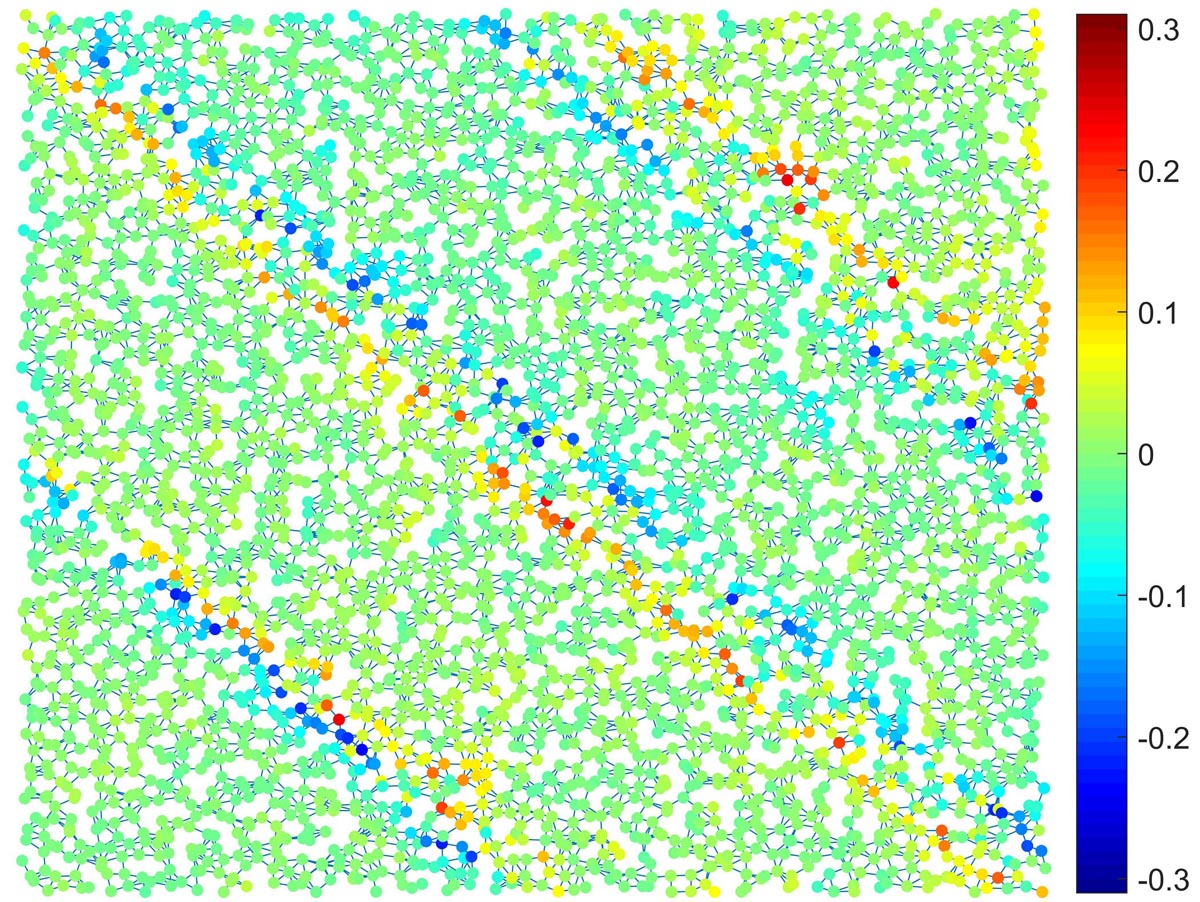}\\
\includegraphics[width=40mm, height=32mm]{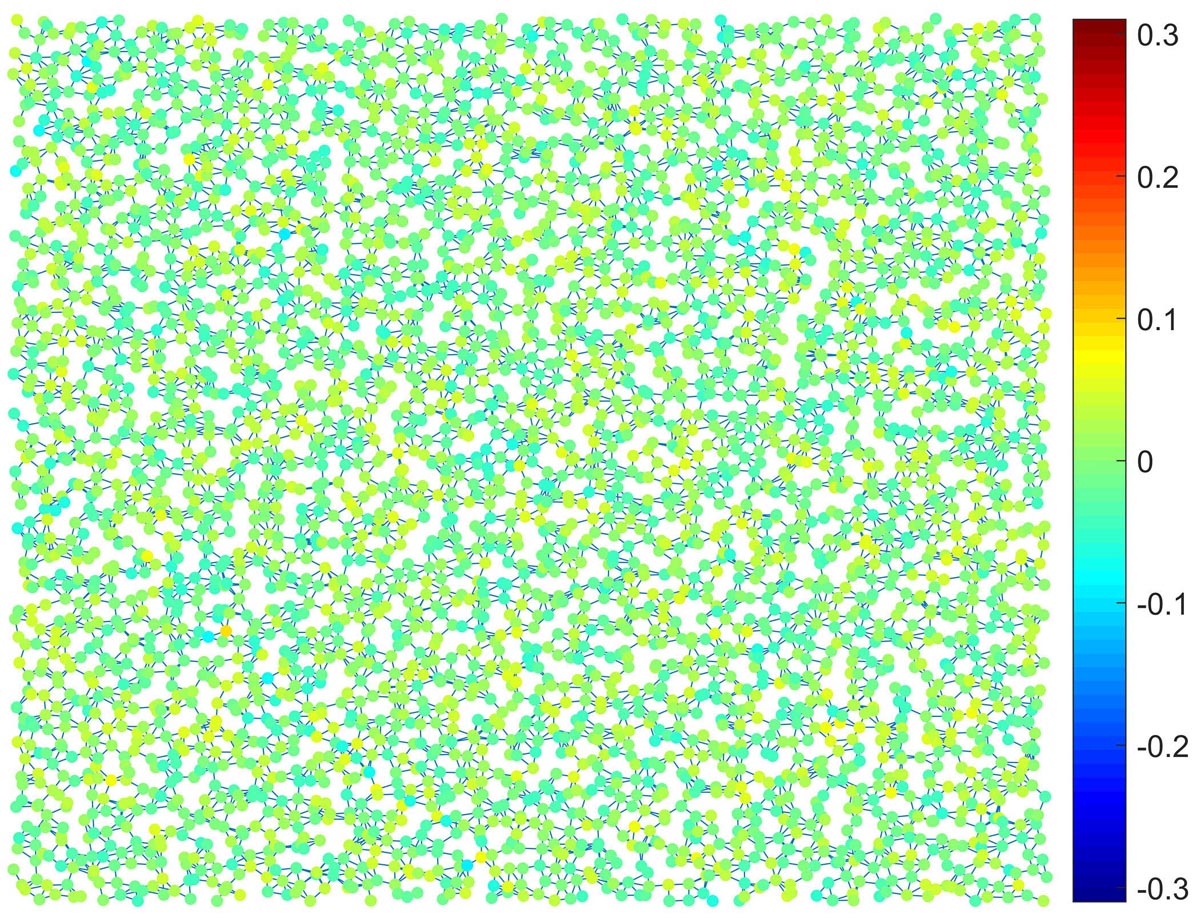}
\includegraphics[width=40mm, height=32mm]{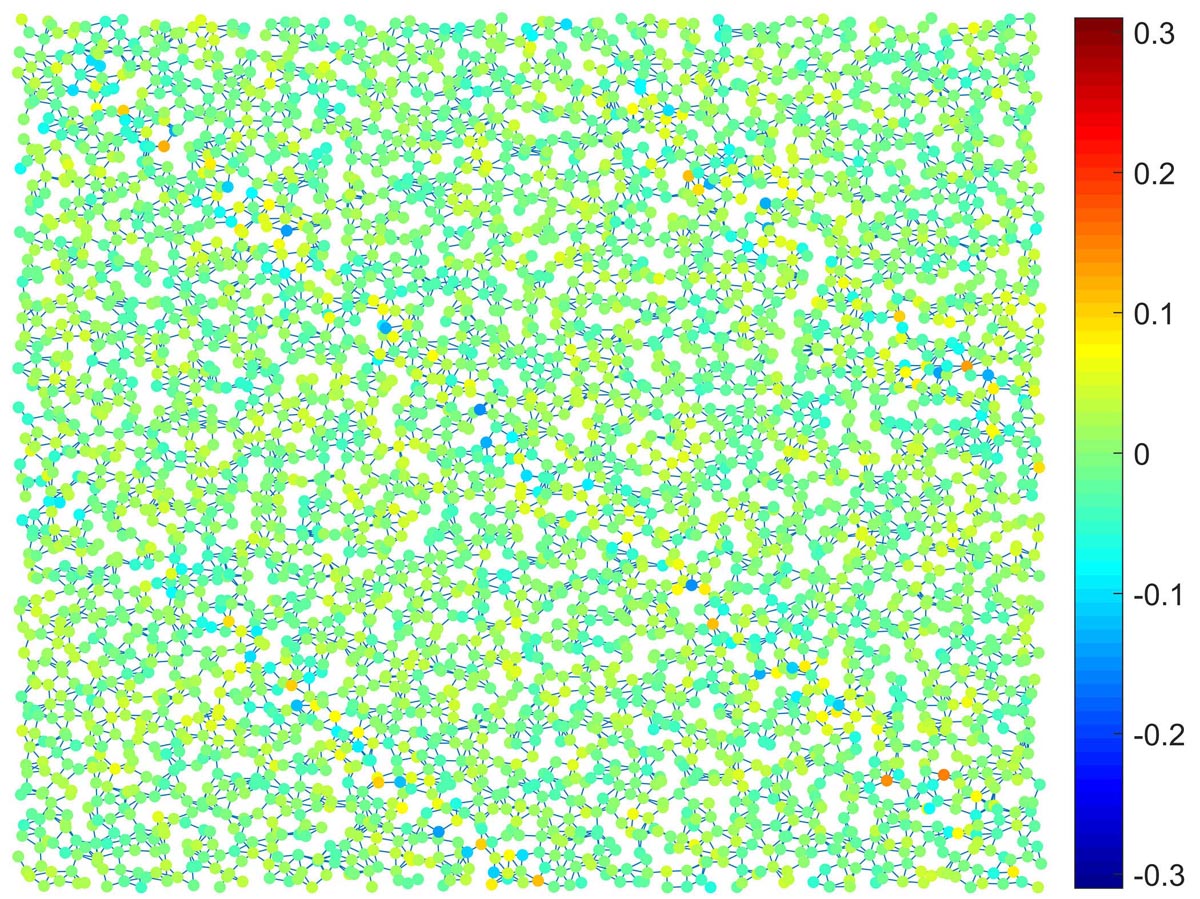}
\includegraphics[width=40mm, height=32mm]{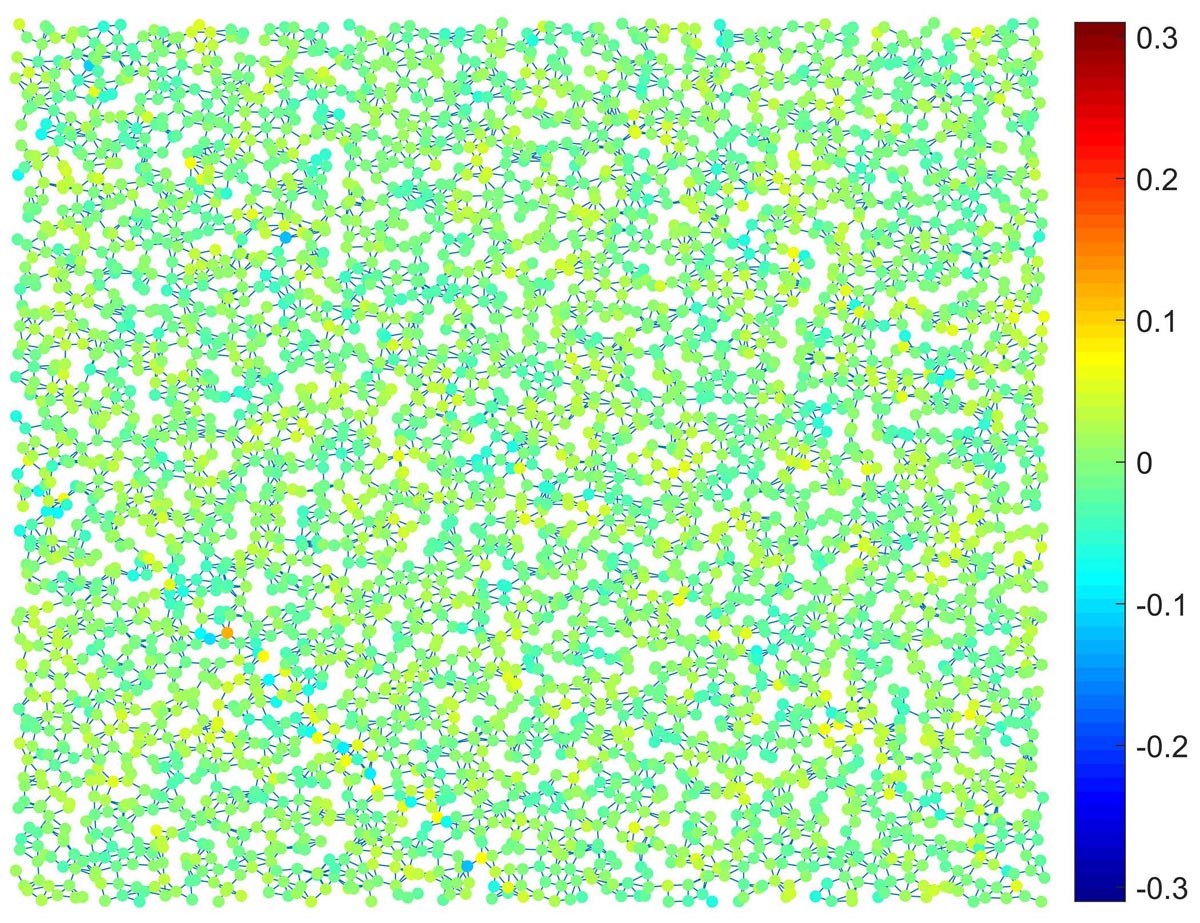}
\includegraphics[width=40mm, height=32mm]{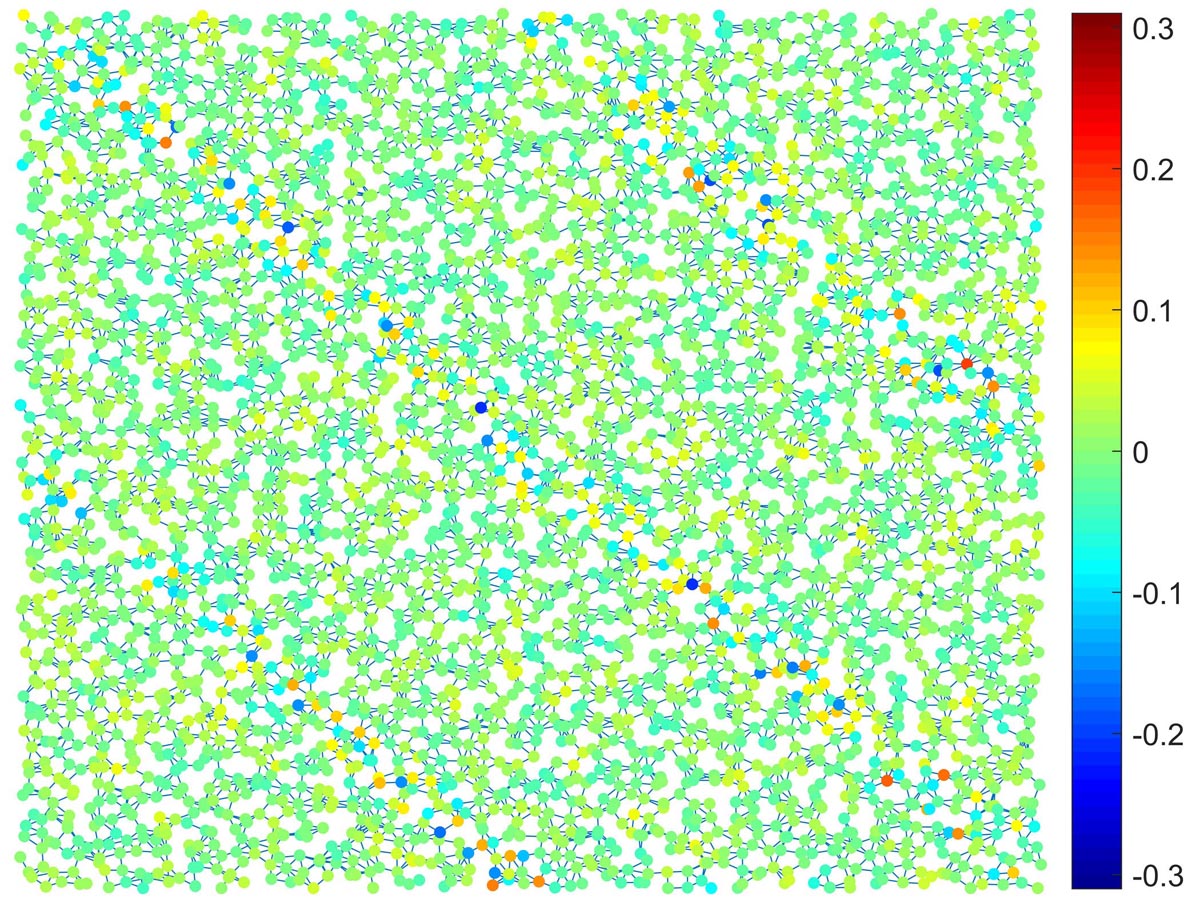}
\caption{Plotted on the top from left to right are the input noise $\pmb \epsilon$,
differences between the original signal ${\bf x}_o$ and the denoised signal $\tilde {\bf x}$ obtained by the graphBior, OSGFB and PRT.
Shown at the bottom from left to right are differences between the original signal ${\bf x}_o$ and the denoised signal $\tilde {\bf x}$ obtained by
 NSGFB-B1, NSGFB-B2, NSGFB-L1 and NSGFB-L2.}
\label{randomgraph_denoise.fig}
\end{figure*}

 Shown in Figure  \ref{randomgraph_denoise.fig} is the input noise $\pmb \epsilon$ with  $\eta=1/16$
 and  differences between the original signal ${\bf x}_o$ and the denoised signal $\tilde {\bf x}$ via the  graphBior,  OSGFB, PRT
 and  spline
 NSGFBs, where a random geometric graph ${\rm RGG}_{4096}$,  original signal ${\bf x}_o$ and  noise ${\pmb \epsilon}$
 are the same as in Tables \ref{Denoise_RRG2.table} and \ref{Denoise_gnninfinity.table}.
 It indicates that
 all denoising techniques have satisfactory performance inside the same strip where the signal has small variation, 
 and  that the spline NSGFBs proposed in this paper
achieve better performance visually on noise suppression
than the other three methods do near the boundary of two adjacency strips where the signal has large variation.

The proposed  NSGFBs 
  can be implemented in a distributed manner
and they  are beneficial to (local) noise suppression 
on graphs of very large scale.
Our simulations  indicate that for random geometric graphs  ${\rm RGG}_N$ with large size $N$ and $1\le p\le  \infty$,
 the output $\ell^p$-SNRs of
spline NSGFBs
 have invisible 
 change for the same input noise level  when
 the graph size $N$  increases.

\appendix

\subsection{Proof of  Proposition \ref{finiteband.pr}}
\label{appendixA}
The first inequality follows from  \eqref{infty.estimate}. Now we prove the second inequality.
 Write ${\bf A}=(a(i,j))_{i,j\in V}$, and define its  Schur norm by
\begin{equation*}\label{schur.def}
\|{\bf A}\|_{{\mathcal S}}=\max
\Big(\sup_{i\in V} \sum_{j\in V} |a(i,j)|,\
\sup_{j\in V} \sum_{i\in V} |a(i,j)|\Big).
\end{equation*}
It is well known that  
 the filter bound $\|{\bf A}\|_{{\mathcal B}_p}, 1\le p\le \infty$, of a graph filter ${\bf A}$
 is dominated by its Schur norm,
\begin{equation}\label{schurbounded}
\|{\bf A}\|_{{\mathcal B}_p}
\le \|{\bf A}\|_{\mathcal S}\ {\rm for \ all}  \  1\le p\le \infty.
\end{equation}
Then it suffices to prove
\begin{equation}\label{finiteband.pr.eq1++}
\|{\bf A}\|_{{\mathcal S}}
\le D_1({\mathcal  G})  (\sigma+1)^d \|{\bf A}\|_{\infty}.
\end{equation}
For any $i\in V$, we obtain
\begin{eqnarray} \label{finiteband.pr.pf.eq1}
\sum_{j\in V} |a(i,j)|  & \hskip-0.08in = &  \hskip-0.08in \sum_{\rho(i,j)\le \sigma} |a(i,j)| \le \|{\bf A}\|_{\infty}\hskip-0.03in \sum_{\rho(i,j)\le \sigma} 1\nonumber\\
& \hskip-0.08in\le &  \hskip-0.08in   D_1({\mathcal  G})  (\sigma+1)^d \|{\bf A}\|_{\infty},
\end{eqnarray} where the second inequality follows from \eqref{countmeasure.pr.eq1}.
Similarly for any $j\in V$, we have
\begin{equation} \label{finiteband.pr.pf.eq2}
\sum_{i\in V} |a(i,j)|  \le D_1({\mathcal  G})  (\sigma+1)^d  \|{\bf A}\|_{\infty}.
\end{equation}
Combining \eqref{finiteband.pr.pf.eq1} and \eqref{finiteband.pr.pf.eq2}  
completes the proof.

\subsection{Proof of Theorem \ref{filterbankpnorm.thm}}
The upper bound estimate \eqref{filterbanknorm.thm.eq2} follows directly from  Proposition \ref{finiteband.pr}
and the observation that
\begin{equation}\label{h.observation}
\|{\bf H}_0\|_{{\mathcal B}_2} +  \|{\bf H}_1\|_{{\mathcal B}_2}\le 2  \|{\bf H}\|_{{\mathcal B}_2}^{1/2}.
\end{equation}
Now we prove the lower bound estimate \eqref{filterbanknorm.thm.eq1}.
Set
\begin{equation}  \label{filterbankpnorm.thm.pf.eq3--}
{\bf B}= {\bf I}- \frac{ {\bf H}}{\|{\bf H}\|_{{\mathcal B}_2}}.\end{equation}
Then  
  ${\bf B}$ has bandwidth $2\sigma$,
\begin{equation}\label{filterbankpnorm.thm.pf.eq3+}
\|{\bf B}\|_{{\mathcal B}_2}\le (\kappa-1)/\kappa, \end{equation}
and
 \begin{equation} \label{filterbankpnorm.thm.pf.eq3++}
 {\bf H}^{-1}= (\|{\bf H}\|_{{\mathcal B}_2})^{-1}\sum_{n=0}^\infty {\bf B}^n.\end{equation}
  Write ${\bf H}^{-1}=(g(i,j))_{i,j\in V}$.  For $\kappa=1$, we have
  \begin{equation}\label{kappa.one}
  {\bf H}^{-1}= (\|{\bf H}\|_{{\mathcal B}_2})^{-1} {\bf I}.
  \end{equation}
  Now we consider the case that $\kappa>1$.
  Set $\theta= \ln (\kappa/(\kappa-1))$, and  for  $i,j\in V$ let $n_0(i,j)$ be the minimal  integer such that $2 n_0(i,j)\ge \rho(i,j)/\sigma$.
   Then
 \begin{eqnarray} \label{filterbankpnorm.thm.pf.eq4}
 \hskip-0.29in
|g(i,j)| & \hskip-0.09in \le & \hskip-0.09in  (\|{\bf H}\|_{{\mathcal B}_2})^{-1}
\sum_{n=n_0(i,j)}^\infty \|{\bf B}^n \|_{\infty}\nonumber\\
& \hskip-0.09in \le & \hskip-0.09in  (\|{\bf H}\|_{{\mathcal B}_2})^{-1}
\sum_{n=n_0(i,j)}^\infty \|{\bf B}\|_{{\mathcal B}_2}^n \nonumber\\
& \hskip-0.09in \le &  \hskip-0.09in (\|{\bf H}\|_{{\mathcal B}_2})^{-1} \kappa (1-\kappa^{-1})^{n_0(i,j)}\nonumber\\
& \hskip-0.09in \le & \hskip-0.09in \|{\bf H}^{-1}\|_{{\mathcal B}_2} \exp \Big( -\frac{\theta}{2\sigma}\rho(i,j)\Big),
\end{eqnarray}
where the first inequality  follows from \eqref{filterbankpnorm.thm.pf.eq3++} and
the observation that ${\bf B}^n $ have bandwidth  $2n \sigma$,
the second one is true by \eqref{infty.estimate}, and the third one  holds by
 \eqref{finiteband.pr.eq1} and  \eqref{filterbankpnorm.thm.pf.eq3+}.

From \eqref{kappa.one} we immediately get
\begin{equation}  \label{filterbankpnorm.thm.pf.eq4+-}
 \|{\bf H}^{-1}\|_{{\mathcal B}_p}=  \|{\bf H}^{-1}\|_{{\mathcal B}_2}
\end{equation}
if $\kappa=1$,
and
by  \eqref{schurbounded} 
and \eqref {filterbankpnorm.thm.pf.eq4}, we have
 \begin{eqnarray}  \label{filterbankpnorm.thm.pf.eq4+}\hskip-0.08in
 \|{\bf H}^{-1}\|_{{\mathcal B}_p}
 & \hskip-0.08in \le  & \hskip-0.08in  \|{\bf H}^{-1}\|_{{\mathcal B}_2}\times
 \nonumber\\
 & \hskip-0.08in   & \hskip-0.08in   
  \sup_{i\in V} \sum_{n=0}^\infty \sum_{ 2n\sigma\le  \rho(i,j)< 2(n+1)\sigma}\exp \Big( \hskip-0.05in -\frac{\theta}{2\sigma}\rho(i,j)\Big)\nonumber\\
   & \hskip-0.08in \le  & \hskip-0.08in  \|{\bf H}^{-1}\|_{{\mathcal B}_2}
  \sup_{i\in V} \sum_{n=0}^\infty e^{-n\theta} \mu\Big(B\big(i, 2(n+1)\sigma-1\big)\Big)\nonumber\\
   & \hskip-0.08in \le  & \hskip-0.08in  (2\sigma)^d D_1({\mathcal G})  \|{\bf H}^{-1}\|_{{\mathcal B}_2}
 \sum_{n=0}^\infty (n+1)^d  (1-\kappa^{-1})^n\nonumber\\
   & \hskip-0.08in \le  & \hskip-0.08in  (2\sigma)^d D_1({\mathcal G}) \|{\bf H}^{-1}\|_{{\mathcal B}_2}
    \Big(\Big( \frac{1}{1-t}\Big)^{(d)}\Big|_{t=1-\kappa^{-1}}\Big)\nonumber \\
 & \hskip-0.08in \le  & \hskip-0.08in   d!  (2\sigma)^d  D_1({\mathcal G}) \kappa^{d+1} \|{\bf H}^{-1}\|_{{\mathcal B}_2}
 \end{eqnarray}
 if $\kappa>1$.
 Then
\begin{eqnarray}
\|{\bf x}\|_p & \hskip-0.08in \le  & \hskip-0.08in \|{\bf H}^{-1} \|_{{\mathcal B}_p} \big(\|{\bf H}_0^T\|_{{\mathcal B}_p} \|{\bf H}_0{\bf x}\|_p+
\|{\bf H}_1^T\|_{{\mathcal B}_p} \|{\bf H}_1{\bf x}\|_p\big)\nonumber\\
& \hskip-0.08in \le & \hskip-0.08in   d!   (2\sigma)^d D_1({\mathcal G}) \kappa^{d+1} \|{\bf H}^{-1}\|_{{\mathcal B}_2}
\nonumber\\
  & &  \times
  \big( \|{\bf H}_0^T\|_{{\mathcal B}_p} \|{\bf H}_0{\bf x}\|_p+
\|{\bf H}_1^T\|_{{\mathcal B}_p} \|{\bf H}_1{\bf x}\|_p\big)
\nonumber\\
& \hskip-0.08in \le & \hskip-0.08in   d!   2^d  (\sigma+1)^{2d} ( D_1({\mathcal G}))^2 \kappa^{d+1} \|{\bf H}^{-1}\|_{{\mathcal B}_2}
\nonumber\\
  & &  \times
  \big( \|{\bf H}_0\|_{{\mathcal B}_2} \|{\bf H}_0{\bf x}\|_p+
\|{\bf H}_1\|_{{\mathcal B}_2}\|{\bf H}_1{\bf x}\|_p\big)\nonumber\\
& \hskip-0.08in \le & \hskip-0.08in
d!   2^{d+1}  (\sigma+1)^{2d} ( D_1({\mathcal G}))^2 \kappa^{d+2}  \|{\bf H}\|_{{\mathcal B}_2}^{-1/2}\nonumber\\
  & & \times \max
  \big( \|{\bf H}_0{\bf x}\|_p,  \|{\bf H}_1{\bf x}\|_p\big), \nonumber\\
  & \hskip-0.08in \le & \hskip-0.08in
d!   2^{d+1}  (\sigma+1)^{2d} ( D_1({\mathcal G}))^2 \kappa^{d+2}  \|{\bf H}\|_{{\mathcal B}_2}^{-1/2}\nonumber\\
  & & \times
  \big( \|{\bf H}_0{\bf x}\|_p^p+  \|{\bf H}_1{\bf x}\|_p^p\big)^{\frac1p}, 
\end{eqnarray}
 where the second inequality follows from  \eqref {filterbankpnorm.thm.pf.eq4+-} and
\eqref{filterbankpnorm.thm.pf.eq4+}, the third holds by  Proposition \ref{finiteband.pr},
and the  fourth one is true by \eqref{h.observation} and \eqref{kappa.def}.
 This proves \eqref{filterbanknorm.thm.eq1} and completes the proof.

 \subsection{Proof of Theorem \ref{bezout.thm}}
  By \eqref{laplacian.decomposition}, \eqref{bezout.thm.eq1}, \eqref{bezout.thm.eq3}
and \eqref{bezout.thm.eq2}, we obtain
\begin{eqnarray*}  \label{bezout.thm.pf.eq1}
 &\hskip-0.08in  &   \hskip-0.08in {\bf G}_0 {\bf H}_0+ {\bf G}_1 {\bf H}_1\nonumber\\
& \hskip-0.08in = &  \hskip-0.08in  Q_0 ({\bf L}_{\mathcal G}^{\rm sym})
 P_0 ({\bf L}_{\mathcal G}^{\rm sym})+  Q_1 ({\bf L}_{\mathcal G}^{\rm sym})  Q_1 ({\bf L}_{\mathcal G}^{\rm sym})\nonumber\\
 & \hskip-0.08in = &  \hskip-0.08in  {\bf U}^T ( Q_0 ({\pmb \Lambda})  P_0 ({\pmb \Lambda})+ Q_1 ({\pmb \Lambda}) P_1 ({\pmb \Lambda})
){\bf U}
  =  {\bf U}^T  {\bf U}={\bf I}.
\end{eqnarray*}
This completes the proof.

 \subsection{Proof of Proposition \ref{bezout.error.pr}}
Set  ${\bf z}_0= {\bf H}_0 {\bf x}$ and ${\bf z}_1={\bf H}_1 {\bf x}$. Then
\begin{eqnarray} \label{bezout.error.pr.pf.eq2}
\|\tilde {\bf x}-{\bf x}\|_p  &\hskip-0.08in  \le & \hskip-0.08in \| {\bf G}_0 ({\bf z}_0- \Psi_0({\bf z}_0))\|_p+
\| {\bf G}_1 ({\bf z}_1- \Psi_1({\bf z}_1))\|_p\nonumber\\
&\hskip-0.08in  \le & \hskip-0.08in (\| {\bf G}_0 \|_{{\mathcal B}_p} + \|{\bf G}_1\|_{{\mathcal B}_p})
\epsilon \nonumber\\
&\hskip-0.08in  \le & \hskip-0.08in D_1({\mathcal G}) (\tilde \sigma+1)^d (\|{\bf G}_0\|_\infty+ \|{\bf G}_1\|_\infty) \epsilon,
\end{eqnarray}
where the first inequality follows from 
 the perfect reconstruction condition \eqref{biorthogonal.condition}
for the NSGFB  constructed in Theorem \ref{bezout.thm}, the second one holds by \eqref{bezout.error.pr.eq1}, and the last estimate is true by Proposition
 \ref{finiteband.pr}.

 \subsection{Proof of Theorem \ref{synthesisdecay.thm}}

 By \eqref{pseudoinverse} and \eqref{filterbankpnorm.thm.pf.eq4}, we have
\begin{eqnarray*}\label{g1.exponential*}
|g_{l, L}(i,j)| & \hskip-0.08in \le & \hskip-0.08in   \|{\bf H}^{-1}\|_{{\mathcal B}_2} \|{\bf H}_l\|_\infty   \sum_{\rho(k, j)\le \sigma}   \exp \Big( -\frac{\theta}{2\sigma}\rho(i,k)\Big)\\
& \hskip-0.08in \le & \hskip-0.08in
D_1({\mathcal G}) \|{\bf H}^{-1}\|_{{\mathcal B}_2} \|{\bf H}_l\|_\infty  (\sigma+1)^{d} \nonumber\\ 
& & \times
\exp \Big( -\frac{\theta}{2\sigma}\rho(i,j)+\frac{\theta}{2}\Big), \ i, j\in V,
\end{eqnarray*}
where $l=0, 1$.  This proves \eqref{g1.exponential}.

\subsection{Proof of Theorem \ref{convergence.thm}}
\label{convergence.section}
Set
${\bf y}^{(m)}= \tilde {\bf x}- {\bf x}^{(m)}$ and write $ {\bf y}^{(m)}=(y^{(m)}(i))_{i\in V}, m\ge 0$.
We claim that
\begin{equation}\label{convergence.thm.pf.eq1}
{\bf y}^{(m)}={\bf H}^{-1} ({\bf H}_0^T \tilde{\bf z}_0^{(m)}+ {\bf H}_1^T \tilde{\bf z}_1^{(m)}),\  m\ge 0.
\end{equation}
The above claim holds for $m=0$, since
$${\bf y}^{(0)}=\tilde {\bf x}=
{\bf H}^{-1} ({\bf H}_0^T \tilde{\bf z}_0+ {\bf H}_1^T \tilde{\bf z}_1) =
{\bf H}^{-1} \big({\bf H}_0^T \tilde{\bf z}_0^{(0)}+ {\bf H}_1^T \tilde{\bf z}_1^{(0)}\big)
$$
by  \eqref{leastsquares.signals} and \eqref{iterativedistributedalgorithm.eq0}.
Inductively for $m\ge 1$, we have
\begin{eqnarray*}
{\bf y}^{(m)} & \hskip-0.08in = & \hskip-0.08in  {\bf y}^{(m-1)}-{\bf v}^{(m)}\\
& \hskip-0.08in = & \hskip-0.08in
{\bf H}^{-1} \big({\bf H}_0^T \tilde{\bf z}_0^{(m-1)}+ {\bf H}_1^T \tilde{\bf z}_1^{(m-1)}\big)- {\bf v}^{(m)}\\
& \hskip-0.08in = & \hskip-0.08in
{\bf H}^{-1} \big({\bf H}_0^T \tilde{\bf z}_0^{(m)} + {\bf H}_1^T \tilde{\bf z}_1^{(m)}\big)
\end{eqnarray*}
where the first and third equalities follow from \eqref{iterativedistributedalgorithm.eqm}
and the second equality holds by the inductive hypothesis.
This completes the proof of Claim \ref{convergence.thm.pf.eq1}.

Write $ (\chi_k^{2r} {\bf H} \chi_k^{2r})^{-1}= (g_k(i,j))_{i,j\in B(k, 2r)}$
and
 \begin{equation}\label{convergence.thm.pf.eq2}
 \chi_k^r (\chi_k^{2r} {\bf H} \chi_k^{2r})^{-1} \chi_k^{2r} {\bf H} (\chi_k^{2r+2\sigma}-\chi_k^{2r})=(\tilde g_k(i,j))_{i,j\in V}, k\in V.\end{equation}
Following the argument used to prove  \eqref{filterbankpnorm.thm.pf.eq4}, we have
\begin{eqnarray} \label{convergence.thm.pf.eq3}
\hskip-0.33in |g_k(i,j)|
& \hskip-0.1in \le & \hskip-0.1in \|{\bf H}^{-1}\|_{{\mathcal B}_2} \exp \Big( -\frac{\theta}{2\sigma}\rho(i,j)\Big)
\end{eqnarray}
for all $i,j\in B(k, 2r)$.  By \eqref{infty.estimate}, \eqref{convergence.thm.pf.eq2} and \eqref{convergence.thm.pf.eq3},
we obtain
\begin{equation}\label{convergence.thm.pf.eq4}
\tilde g_k(i,j)=0
\end{equation}
where  either $i\not \in B(k, r)$ or $j\not\in B(k, 2r+2\sigma)\backslash B(k, 2r)$,
and
\begin{eqnarray}\label{convergence.thm.pf.eq5}
|\tilde g_k(i,j)| & \hskip-0.08in \le & \hskip-0.08in  \|{\bf H}^{-1}\|_{{\mathcal B}_2} \|{\bf H}\|_\infty
\sum_{l\in B(j,2\sigma)}  \exp \Big( -\frac{\theta}{2\sigma}\rho(i,l)\Big)\nonumber\\
& \hskip-0.08in \le & \hskip-0.08in  D_1({\mathcal G}) (2\sigma+1)^d \kappa   \exp \Big( -\frac{\theta }{2\sigma}  r+\theta\Big)
\end{eqnarray}
where  $i\in B(k,r)$ and $j\in B(k, 2r+2\sigma)\backslash B(k, 2r)$.

Write
 ${\bf v}^{(m)}_k= (v^{(m)}_k(i))_{i\in V}, m\ge 1, k\in V$. By \eqref{iterativedistributedalgorithm.eqm},
 \eqref{convergence.thm.pf.eq1}, we have
  \begin{eqnarray*} \chi_k^r({\bf v}^{(m)}_k-{\bf y}^{(m-1)}) & \hskip-0.08in = & \hskip-0.08in
 \chi_k^r (\chi_k^{2r} {\bf H} \chi_k^{2r})^{-1} \chi_k^{2r} \nonumber\\
 & & \times {\bf H} (\chi_k^{2r+2\sigma}-\chi_k^{2r}) {\bf y}^{(m-1)}. \end{eqnarray*}
Combining the above equation with
 \eqref{convergence.thm.pf.eq4} and \eqref{convergence.thm.pf.eq5}, we get
 \begin{eqnarray}\label{convergence.thm.pf.eq6}
& \hskip-0.08in  & \hskip-0.08in   |v^{(m)}_k(i)-y^{(m-1)}(i)| = \Big|\sum_{j\in B(k, 2r+2\sigma)}
 \tilde g_k(i,j) y^{(m-1)}(j)\Big|\nonumber\\
  & \hskip-0.08in \le  & \hskip-0.08in
 D_1({\mathcal G}) (2\sigma+1)^d \kappa   \exp \Big( -\frac{\theta }{2\sigma}  r+\theta\Big)
 \nonumber\\
   & \hskip-0.08in & \hskip-0.08in \times  \Big (\sum_{j\in B(i, 3r+2\sigma)}
 | y^{(m-1)}(j)|\Big), \ \ i\in B(k, r).
 \end{eqnarray}
This together with \eqref{iterativedistributedalgorithm.eqm} implies that
\begin{eqnarray}
  |y^{(m)}(i)| & \hskip-0.08in =  & \hskip-0.08in   |v^{(m)}(i)-y^{(m-1)}(i)|\nonumber  \\
   & \hskip-0.08in \le  & \hskip-0.08in \frac{1}{\mu(B(i,r))}
\sum_{k\in B(i,r)} |v^{(m)}_k(i)-y^{(m-1)}(i)|\nonumber\\
  & \hskip-0.08in \le  & \hskip-0.08in
 D_1({\mathcal G}) (2\sigma+1)^d \kappa   \exp \Big( -\frac{\theta }{2\sigma}  r+\theta\Big)
 \nonumber\\
   & \hskip-0.08in & \hskip-0.08in \times  \Big (\sum_{j\in B(i, 3r+2\sigma)}
 | y^{(m-1)}(j)|\Big)
\end{eqnarray}
for all $i\in V$ and $m\ge 1$.
Using the above componentwise estimate, we obtain
\begin{equation}
\|{\bf y}^{(m+1)}\|_p\le \delta_{r, \sigma} \|{\bf y}^{(m)}\|_p, \ m\ge 0.
\end{equation}
Iteratively applying the above estimate proves \eqref{convergence.thm.eq3}.

\end{document}